# Beyond the Elementary Representations of Program Invariants over Algebraic Data Types

## Extended Version

Yurii Kostyukov
y.kostyukov@2015.spbu.ru
Saint Petersburg State University
JetBrains Research
Russia

Dmitry Mordvinov
dmitry.mordvinov@jetbrains.com
Saint Petersburg State University
JetBrains Research
Russia

Grigory Fedyukovich
grigory@cs.fsu.edu
Florida State University
USA

## Abstract

First-order logic is a natural way of expressing properties of computation. It is traditionally used in various program logics for expressing the correctness properties and certificates. Although such representations are expressive for some theories, they fail to express many interesting properties of algebraic data types (ADTs). In this paper, we explore three different approaches to represent program invariants of ADT-manipulating programs: tree automata, and first-order formulas with or without size constraints. We compare the expressive power of these representations and prove the negative definability of both first-order representations using the pumping lemmas. We present an approach to automatically infer program invariants of ADT-manipulating programs by a reduction to a finite model finder. The implementation called RInGen has been evaluated against state-of-the-art invariant synthesizers and has been experimentally shown to be competitive. In particular, program invariants represented by automata are capable of expressing more complex properties of computation and their automatic construction is often less expensive.

## 1 Introduction

Specifying and proving properties of programs is traditionally achieved with the help of first-order logic (FOL). It is widely used in various techniques for verification, from Floyd-Hoare logic [24, 30] to constrained Horn clauses (CHC) [7] and refinement types [56]. The language of FOL allows to describe the desired properties precisely and make the verification technology accessible to the end user. Similarly, verification proofs, such as inductive invariants, procedure summaries, or ranking functions are produced and returned to the user also in FOL, thus facilitating the explainability of a program and its behaviors.

In this paper, we demonstrate theoretically and practically that the class of solutions traditionally considered by state-of-the-art FOL-based tools is not wide enough to fulfill the expectations from automated verification. Despite offering a high degree of expressiveness, decision procedures and constraint solvers are sensitive to particular FOL-fragments and have an emerging need for algorithmic improvement. Algebraic Data Types (ADT) enjoy a variety of decision procedures [4, 45, 48, 53] and Craig interpolation algorithms [31, 34], but still many practical tasks cannot be solved by state-of-the-art solvers for Satisfiability Modulo Theory (SMT) such as Z3, CVC4 [2], and Princess [51].

Furthermore, with the recent growth of the use of SMT solvers, it is often tempting to formulate verification conditions using the combination of different theories, e.g., as in [21]. Verification conditions could be expressed using the combination of ADT and the theory of Equality and Uninterpreted Functions (EUF). Although SMT solvers claim to support EUF, in reality the proof search process often hangs back attempting to conduct structural induction and discovering helper lemmas [57].

In this paper, we introduce a new *automata-based* class of representations of inductive invariants. The basic idea is to find a finite model of the verification condition and convert this model into a finite automaton. Instead of representing program states, finite models describe the states of the tree automaton via a known correspondence between tree automata and finite models. The resulting representations of invariants are *regular* in a sense that they can "scan" the ADT term to the unbounded depth, which cannot be reached by the representations by first-order formulas (called *elementary* throughout the paper).

Our first contribution is the demonstration that regular invariants of ADT-manipulating programs could be constructed from finite models of the verification condition. Intuitively, the invariant generation problem can be reduced to the satisfiability problem of a formula constructed from the FOL-encoding of the program with pre- and post-conditions where uninterpreted symbols are used instead of ADT constructors. Although becoming an over-approximation of the original verification condition, it can be handled by existing finite model finders, such as MACE4 [44], FINDER [52], PARADOX [13], or CVC4 [50]. If satisfiable, the detected model is used to construct regular solutions of the original problem.

We have investigated the expressiveness of three different representations of program invariants: 1) state-of-the-art elementary representations, 2) first-order representations with size constraints, and 3) introduced in this paper regular representations. Our second contribution is the theoretical result on the negative definability of both first-order representations studies in the paper. Knowing about the undefinability lets us understand why SMT-based techniques like Z3 [36] or ELDARICA [32] diverge on correct programs.

We have formulated two *pumping lemmas* for two first-order representations. The concept of pumping lemma arises in the universe of formal languages in connection with finite automata and context-free languages. Pumping lemmas state that for all languages in some class (e.g., regular languages, context-free languages), any big enough element can be "pumped", i.e., get an unbounded increase in some of its parts, and still generate to some elements of the language. Pumping lemmas are useful for proving negative definability: we suppose that a language belongs to some class, then apply a pumping lemma for that class and get some "pumped" element, which cannot be in the language, and by contradiction we deduce that the language does belong to that class.

We have implemented a tool called RInGen for automated inference of the regular invariants of ADT-manipulating programs and evaluated it against state-of-the-art inductive invariant generators that support ADT, namely Z3/SPACER [36], ELDARICA [32], and VERIMAP-IDDT [16]. It managed to find non-trivial invariants of various problems, including the inhabitance checking for simply typed lambda calculus (STLC).

## 2 Motivating Example

In this section, we demonstrate the phenomenon of *inexpressiveness* of a first-order language, in which program invariants can be represented. For example, the following program asserts that there are no two consecutive Peano numbers that are both even.

**Example 1** (*Even*)**.**

```
Nat ::= Z | S Nat
fun even(x : Nat) : bool =
    match x with
    | Z -> true
    | S Z -> false
    | S (S x') -> even(x')
assert ¬(∃x : Nat, even(x) ∧ even(S(x)))
```

This assertion holds, and the program is safe. One standard way to prove it is to discover a *safe inductive invariant*. The invariant could be represented by a first-order formula $even(x)$, satisfying the following logical constraints, which are called the *verification conditions* of the program in the form of constrained Horn clauses (CHC):

$$even(Z) \land \forall x.\bigl(even(x) \to even(S(S(x)))\bigr) \land$$
$$\forall x.\bigl(even(x) \land even(S(x)) \to \bot\bigr)$$

The first-order language of the Nat datatype is the language of successor arithmetic, i.e., the language, that allows for expressing only the first-order combination of arithmetic constraints of the form $x = c$ and $x = y+c$, where $x$ and $y$ are variables, and $c \in \mathbb{N}$. Thus, the successor arithmetic allows to define only finite and co-finite relations (i.e., relations with the finite complement) [19]. The only possible interpretation of *even* satisfying these CHCs is a relation $\{S^{2n}(Z) \mid n \geq 0\}$, which is not expressible in the first-order language of the Nat datatype. However, it could be represented by the automaton which moves to state $s_0$ for $Z$ and flips the state from $s_0$ to $s_1$ and vice versa for $S$:

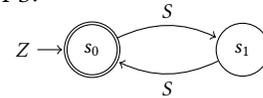

## 3 Preliminaries

**Many-sorted logic.** A many-sorted first-order signature with equality is a tuple $\Sigma = \langle \Sigma_S, \Sigma_F, \Sigma_P \rangle$, where $\Sigma_S$ is a set of sorts, $\Sigma_F$ is a set of function symbols, $\Sigma_P$ is a set of predicate symbols, containing equality symbol $=_\sigma$ for each sort $\sigma$. Each function symbol $f \in \Sigma_F$ has an associated arity of the form $\sigma_1 \times \cdots \times \sigma_n \to \sigma$, where $\sigma_1, \ldots, \sigma_n, \sigma \in \Sigma_S$, and each predicate symbol $p \in \Sigma_P$ has an associated arity of the form $\sigma_1 \times \cdots \times \sigma_n$. Variables are associated with a sort as well. We use the usual definition of first-order terms with sort $\sigma$, ground terms, formulas, and sentences.

A many-sorted structure $\mathcal{M}$ for a signature $\Sigma$ consists of non-empty domains $|\mathcal{M}|_\sigma$ for each sort $\sigma \in \Sigma_S$. For each function symbol $f$ with arity $\sigma_1 \times \cdots \times \sigma_n \to \sigma$, it associates an interpretation $M(f) : |\mathcal{M}|_{\sigma_1} \times \cdots \times |\mathcal{M}|_{\sigma_n} \to |\mathcal{M}|_\sigma$, and for each predicate symbol $p$ with arity $\sigma_1 \times \cdots \times \sigma_n$ it associated an interpretation $M(p) \subseteq |\mathcal{M}|_{\sigma_1} \times \cdots \times |\mathcal{M}|_{\sigma_n}$. For each ground term $t$ with sort $\sigma$, we define an interpretation $\mathcal{M}[\![t]\!] \in |\mathcal{M}|_\sigma$ in a natural way. We call a structure finite if the domain of every sort is finite; otherwise, we call it infinite.

We assume the usual definition of a satisfaction of a sentence $\varphi$ by $\mathcal{M}$, denoted $\mathcal{M} \models \varphi$. If $\varphi$ is a formula, then we write $\varphi(x_1, \ldots, x_n)$ to emphasize that all free variables of $\varphi$ are among $\{x_1, \ldots, x_n\}$. In this case, we denote the satisfiability $\mathcal{M} \models \varphi(a_1, \ldots, a_n)$ by $\mathcal{M}$ with free variables evaluated to elements $a_1, \ldots, a_n$ of the appropriate domains. The universal closure of a formula $\varphi(x_1, \ldots, x_n)$, denoted $\forall \varphi$, is the sentence $\forall x_1 \ldots \forall x_n.\varphi$. If $\varphi$ has free variables, we define $\mathcal{M} \models \varphi$ to mean $\mathcal{M} \models \forall \varphi$.

A **Herbrand universe** for a sort $\sigma$ is a set of ground terms with sort $\sigma$. If a Herbrand universe for a sort $\sigma$ is infinite, we call $\sigma$ an infinite sort. We say that $\mathcal{H}$ is a *Herbrand structure* $\mathcal{H}$ for a signature $\Sigma$ if it associates the Herbrand universe $|\mathcal{H}|_\sigma$ to each sort $\sigma$ of $\Sigma$ as the domain and interprets every function symbol with itself, i.e., $\mathcal{H}(f)(t_1, \ldots, t_n) = f(t_1, \ldots, t_n)$ for all ground terms $t_i$ with the appropriate sort. Thus, there is a family of Herbrand structures for one signature $\Sigma$ with identical domains and interpretations of function symbols, but with various interpretations of predicate symbols. Every Herbrand structure $\mathcal{H}$ interprets each ground term $t$ with itself, i.e., $\mathcal{H}[\![t]\!] = t$.

**Assertion language.** An algebraic data type (ADT) is a tuple $\langle C, \sigma \rangle$, where $\sigma$ is a sort and $C$ is a set of uninterpreted function symbols (called constructors), such that each $f \in C$ has a sort $\sigma_1 \times \cdots \times \sigma_n \to \sigma$ for some sorts $\sigma_1, \ldots, \sigma_n$.

In what follows, we fix a set of ADTs $\langle C_1, \sigma_1 \rangle, \ldots, \langle C_n, \sigma_n \rangle$ with $\sigma_i \neq \sigma_j$ and $C_i \cap C_j = \emptyset$ for $i \neq j$. We define the signature[1] $\Sigma = \langle \Sigma_S, \Sigma_F, \Sigma_P \rangle$, where $\Sigma_S = \{\sigma_1, \ldots, \sigma_n\}$, $\Sigma_F = C_1 \cup \cdots \cup C_n$, and $\Sigma_P = \{=_{\sigma_1}, \ldots, =_{\sigma_n}\}$. For brevity, we omit the sorts from the equality symbols. We refer to the first-order language defined by $\Sigma$ to as an *assertion language* $\mathcal{L}$.

Because $\Sigma$ has no predicate symbols except the equality symbols (which have fixed interpretations within every structure), then there is a unique Herbrand structure $\mathcal{H}$ for $\Sigma$. We say that a sentence (a formula) $\varphi$ in an assertion language is *satisfiable modulo theory* of ADTs, iff $\mathcal{H} \models \varphi$.

**Constrained Horn Clauses.** We refer to a finite set of predicate symbols $\mathcal{R} = \{P_1, \ldots, P_n\}$ with sorts from $\Sigma$ to as *uninterpreted* symbols.

**Definition 1.** A constrained Horn clause (CHC) $C$ is a $\Sigma \cup \mathcal{R}$-formula of the form:
$$\varphi \land R_1(\overline{t}_1) \land \ldots \land R_m(\overline{t}_m) \to H$$
where $\varphi$ is a formula in the assertion language, called a *constraint*; $R_i \in \mathcal{R}$; $\overline{t}_i$ is a tuple of terms; and $H$, called a *head*, is either $\bot$, or an atomic formula $R(\overline{t})$ for some $R \in \mathcal{R}$.

If $H = \bot$, we say that $C$ is a *query clause*, otherwise we call $C$ a *definite clause*. The premise of the implication $\varphi \land R_1(\overline{t}_1) \land \ldots \land R_m(\overline{t}_m)$ is called a *body* of $C$.

A CHC system $\mathcal{S}$ is a finite set of CHCs.

**Satisfiability of CHCs.** Let $\overline{X} = \langle X_1, \ldots, X_n \rangle$ be a tuple of relations, such that if $P_i$ has sort $\sigma_1 \times \ldots \times \sigma_m$, then $X_i \subseteq |\mathcal{H}|_{\sigma_1} \times \ldots \times |\mathcal{H}|_{\sigma_m}$. To simplify the notation, we denote the expansion $\mathcal{H}\{P_1 \mapsto X_1, \ldots, P_n \mapsto X_n\}$ by $\langle \mathcal{H}, X_1, \ldots, X_n \rangle$, or simply by $\langle \mathcal{H}, \overline{X} \rangle$. We say that a system of CHCs $\mathcal{S}$ is *satisfiable modulo theory* of ADTs, if there exists a tuple of relations $\overline{X}$, such that $\langle \mathcal{H}, \overline{X} \rangle \models C$ for all $C \in \mathcal{S}$.

---
[1] For simplicity, we omit the selectors and testers from the signature because they do not increase the expressiveness of the assertion language.

For example, the system of CHCs from Example 1 is satisfied by interpreting *even* with the relation
$X = \{Z, S(S(Z)), S(S(S(S(Z)))), \ldots\} = \{S^{2n}(Z) \mid n \geq 0\}$.

It is well known that CHCs provide a first-order match for a variety of program logics, including Floyd-Hoare logic for imperative programs and refinement types for higher-order functional programs. Thus, we assume that for every recursive program over ADTs there is a system of CHCs, such that the program is safe iff the system is satisfiable. In the rest of the paper, we use CHCs as a means of expressing verification conditions of programs.

**Definability.** A *representation class* is a function $\mathcal{C}$ mapping every tuple $\langle \sigma_1, \ldots, \sigma_n \rangle \in \Sigma_S^n$ for every $n \in \mathbb{N}$ to some class of languages $\mathcal{C}(\sigma_1, \ldots, \sigma_n) \subseteq 2^{|\mathcal{M}|_{\sigma_1} \times \cdots \times |\mathcal{M}|_{\sigma_n}}$. We say that a relation $X \subseteq |\mathcal{M}|_{\sigma_1} \times \ldots \times |\mathcal{M}|_{\sigma_n}$ is *definable* in a representation class $\mathcal{C}$ if $X \in \mathcal{C}(\sigma_1, \ldots, \sigma_n)$. We say that a Herbrand structure $\mathcal{H}$ is definable in $\mathcal{C}$ (or $\mathcal{C}$-definable) if for every predicate symbol $p \in \Sigma_P$ with arity $\sigma_1 \times \cdots \times \sigma_n$, interpretation $\mathcal{H}[\![p]\!]$ belongs to $\mathcal{C}(\sigma_1, \ldots, \sigma_n)$.

**Finite Tree Automata.** In order to define regular representations, we introduce *deterministic finite tree automata* (DFTA). Let $\Sigma = \langle \cdot, \Sigma_F, \cdot \rangle$ be a fixed many-sorted signature.

**Definition 2** (cf. [14])**.** A *deterministic finite tree n-automaton* over $\Sigma_F$ is a quadruple $(S, \Sigma_F, S_F, \Delta)$, where $S$ is a finite set of states, $S_F \subseteq S^n$ is a set of final states, $\Delta$ is a transition relation with rules of the form:
$$f(s_1, \ldots, s_m) \to s,$$
where $f \in \Sigma_F$, $ar(f) = m$ and $s, s_1, \ldots, s_m \in S$, and there are no two rules in $\Delta$ with the same left-hand side.

**Definition 3.** A tuple of ground terms $\langle t_1, \ldots, t_n \rangle$ is *accepted* by $n$-automaton $A = (S, \Sigma_F, S_F, \Delta)$ iff $\langle A[t_1], \ldots, A[t_n] \rangle \in S_F$, where
$$A[f(t_1, \ldots, t_m)] \stackrel{\text{def}}{=} \begin{cases} s, & \text{if } (f(A[t_1], \ldots, A[t_m]) \to s) \in \Delta, \\ \bot, & \text{otherwise.} \end{cases}$$

**Example 2.** Let $\Sigma = \langle Prop, \{\land, \lor, \to, \top, \bot\}, \emptyset \rangle$ be a signature of propositional logic. The automaton $A$ accepts true propositional formulas without variables:
$$A = (\{q_0, q_1\}, \Sigma_F, \{q_1\}, \Delta),$$
where the $\Delta$ is defined as follows:
$$\bot \mapsto q_0 \qquad q_0 \lor q_0 \mapsto q_0$$
$$\top \mapsto q_1 \qquad * \lor * \mapsto q_1$$
$$q_1 \land q_1 \mapsto q_1 \qquad q_1 \to q_0 \mapsto q_0$$
$$* \land * \mapsto q_0 \qquad * \to * \mapsto q_1.$$

**Regular Herbrand Models.** Let $\mathcal{H}$ be a Herbrand structure for a signature $\langle \cdot, \Sigma_F, \cdot \rangle$. We say that $n$-automaton $A$ over $\Sigma_F$

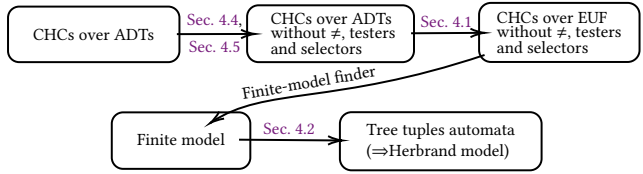

**Figure 1.** Obtaining regular model of a CHC system over ADTs.

*represents* a relation $X \subseteq |\mathcal{H}|_{\sigma_1} \times \ldots \times |\mathcal{H}|_{\sigma_n}$ iff $X = \{\langle a_1, \ldots, a_n \rangle \mid \langle a_1, \ldots, a_n \rangle \text{ is accepted by } A, a_i \in |\mathcal{H}|_{\sigma_i}\}$. If there is a DFTA representing $X$, we call $X$ *regular*. We denote the class of regular relations by REG. A structure $\mathcal{H}$ is regular if it is REG-definable.

## 4 Automated Inference of Regular Invariants

In this section, we show how to obtain regular models of CHCs over ADTs using a finite model finder, e.g., [13, 44, 50, 52]. Figure 1 gives the main steps of this process: Given a system of CHCs, we first rewrite it into a formula over uninterpreted function symbols by eliminating all disequalities, testers, and selectors from the clause bodies. Then we reduce the satisfiability modulo theory of ADTs to satisfiability modulo EUF and apply a finite model finder to build a finite model of the reduced verification conditions. Finally, using the correspondence between finite models and tree automata, we get the automaton representing the safe inductive invariant.

### 4.1 Translation to EUF

Recall that by definition, we call the system of CHCs over ADTs satisfiable if every clause is satisfied in some expansion of the Herbrand structure. The main insight is that this satisfiability problem can be reduced to checking the satisfiability of a formula over uninterpreted symbols in a usual first-order sense.

Informally, given a system of CHCs, we obtain another system by replacing all ADT constructors in all CHCs with uninterpreted function symbols. This allows the interpretations of constructors to violate the ADT axioms (distinctiveness, injectivity, exhaustiveness, etc.). The system with uninterpreted symbols is either satisfiable or unsatisfiable in the usual first-order sense. In the former case every clause is satisfied by some structure $\mathcal{M}$. It could be used to extract the interpretations of uninterpreted symbols in the Herbrand structure $\mathcal{H}$ which satisfy the original system over $\mathcal{H}$.

For instance, for the system of CHCs in Example 1, we check the satisfiability of the following formula:

$$\forall x.(x = Z \to even(x)) \land$$
$$\forall x, y.(x = S(S(y)) \land even(y) \to even(x)) \land$$
$$\forall x, y.(even(x) \land even(y) \land y = S(x) \to \bot).$$

The formula is satisfied by the following finite model $\mathcal{M}$:

$$|\mathcal{M}|_{Nat} = \{0, 1\} \qquad \mathcal{M}(Z) = 0$$
$$\mathcal{M}(even) = \{0\} \qquad \mathcal{M}(S)(x) = 1 - x.$$

### 4.2 Finite Models To Tree Tuples Automata

A procedure for constructing tree tuples automata (and, hence, regular models) from finite models follows immediately from the construction of an isomorphism between finite models and tree automata [41].

Given a finite structure $\mathcal{M}$, we build an automaton $\mathcal{A}_P = (|\mathcal{M}|, \Sigma_F, \mathcal{M}(P), \tau)$ for every predicate symbol $P \in \Sigma_P$. A shared set of transitions $\tau$ is defined as follows: for each $f \in \Sigma_F$ with arity $\sigma_1 \times \ldots \times \sigma_n \mapsto \sigma$, for each $x_i \in |\mathcal{M}|_{\sigma_i}$, $\tau(f(x_1, \ldots, x_n)) = \mathcal{M}(f)(x_1, \ldots, x_n)$. For example, $\mathcal{A}_{even}$ is isomorphic to one shown in Example 1.

**Theorem 1.** *For the constructed automaton*
$$\mathcal{A}_P = (S, \Sigma_F, S_F, \tau),$$
$$L(\mathcal{A}_P) = \{\langle t_1, \ldots, t_n \rangle \mid \langle \mathcal{M}[\![t_1]\!], \ldots, \mathcal{M}[\![t_n]\!]\rangle \in \mathcal{M}(P)\}.$$

*Proof.* The proof is straightforward from the fact that $\mathcal{A}_P$ reflects checking the satisfiability in $\mathcal{M}$. □

In practice, this means that CHCs over ADTs could be automatically solved by *finite model finders*, such as MACE4 [44], FINDER [52], PARADOX [13], or CVC4 in a special mode [50]: if a *finite* model (in the usual first-order sense) is found, then there exists a *regular Herbrand* model of the CHC system. In Sec. 8, we evaluate our implementation that uses CVC4 against state-of-art CHC solvers.

### 4.3 Herbrand Models Without Equality

With the correspondence between finite models and tree automata in hand, it remains to show that the Herbrand model induced by the constructed tree automaton is a model of the original CHC system. In this subsection we show that it is straightforward when the system has no disequality constraints, otherwise some additional steps should be done.

Under assumption that the signature $\Sigma$ of the assertion language does not have the equality symbol, there are no predicate symbols at all. Thus we may assume that every constraint in every CHC is $\top$. For instance, the above example could be rewritten to:

$$\top \to even(Z)$$
$$even(x) \to even(S(S(x)))$$
$$even(x) \land even(S(x)) \to \bot.$$

**Lemma 2.** *Suppose that a CHC system $\mathcal{S}$ over uninterpreted symbols $\mathcal{R} = \{P_1, \ldots, P_k\}$ with no constraints is satisfied by some first-order structure $\mathcal{M}$, i.e., $\mathcal{M} \models C$ for all $C \in \mathcal{S}$. Let*
$$X_i \stackrel{\text{def}}{=} \{\langle t_1, \ldots, t_n \rangle \mid \mathcal{M}[\![t_1]\!], \ldots, \mathcal{M}[\![t_n]\!] \in \mathcal{M}(P_i)\}.$$
*Then $\langle \mathcal{H}, X_1, \ldots, X_k \rangle$ is the Herbrand model of $\mathcal{S}$.*

*Proof.* As clause bodies have no constraints, each CHC is of the form $C \equiv R_1(\bar{t}_1) \wedge \ldots \wedge R_m(\bar{t}_m) \to H$. Then by definition
$$\langle \mathcal{H}, X_1, \ldots, X_k \rangle \models C \iff \mathcal{M} \models C,$$
so every clause in $\mathcal{S}$ is satisfied by $\langle \mathcal{H}, X_1, \ldots, X_k \rangle$. □

For Example 1, $X \stackrel{\text{def}}{=} \{t \mid \mathcal{M}[\![t]\!] = 0\} = \{S^{2n}(Z) \mid n \geq 0\}$ indeed satisfies the system.

### 4.4 Herbrand Models With Equality

In the presence of the equality symbol, which has the predefined semantics, a finite model finder searches for a model in a completely free domain, thus breaking the regular model. Consider the system consisting of the only CHC:
$$Z \neq S(Z) \to \bot.$$
This system is unsatisfiable because $\mathcal{H} \models Z \neq S(Z)$. But in a usual first-order sense, i.e., if we treat $Z$ and $S$ as uninterpreted functions, this CHC is satisfiable, e.g., as follows:
$$|\mathcal{M}|_{nat} = \{0\} \qquad \mathcal{M}(Z) = \mathcal{M}(S)(*) = 0$$

In general, every clause with a disequality constraint in the premise may be satisfied by falsifying its premise, e.g., by picking a sort with the cardinality one.

We propose the following way of attacking this problem. For every ADT $\langle C, \sigma \rangle$, we introduce a fresh uninterpreted symbol $diseq_\sigma$ and define $\mathcal{R}' \stackrel{\text{def}}{=} \mathcal{R} \cup \{diseq_\sigma \mid \sigma \in \Sigma_S\}$. Our process proceeds to constructing another system of CHCs $\mathcal{S}'$ over $\mathcal{R}'$. Without loss of generality, we may assume that the constraint of each clause $C \in \mathcal{S}$ is in the Negation Normal Form (NNF). Let $C'$ be a clause with every literal of the form $\neg(t =_\sigma u)$ in the constraint (which we refer to as *disequality constraint*) substituted with the atomic formula $diseq_\sigma(t, u)$. For every clause $C \in \mathcal{S}$, we add $C'$ into $\mathcal{S}'$. Finally, for every ADT $\langle C, \sigma \rangle$, we add the following rules for $diseq_\sigma$ to $\mathcal{S}'$:

- for all distinct $c, c'$ of sort $\sigma$:
$$\top \to diseq_\sigma(c(\bar{x}), c'(\bar{x}'))$$

- for all constructors $c$ of sort $\sigma$, $x$ and $y$ of sort $\sigma'$, and $i$:
$$diseq_{\sigma'}(x, y) \to diseq_\sigma(c(\ldots, \underbrace{x}_{i\text{-th position}}, \ldots), c(\ldots, \underbrace{y}_{i\text{-th position}}, \ldots))$$

It is well-known that universal CHCs admit the least model, which is the denotational semantics of the program modeled by the CHCs, i.e., the least fixed point of the step operator. Thus, the following fact is trivial.

**Lemma 3.** *The rules of $diseq_\sigma$ have the least model over $\mathcal{H}$, which interprets $diseq_\sigma$ by a relation $\mathcal{D}_\sigma$, defined as $\mathcal{D}_\sigma \stackrel{\text{def}}{=} \{(x, y) \in |\mathcal{H}|_\sigma^2 \mid x \neq y\}$ for each sort $\sigma$ in $\Sigma_S$.*

As a corollary of this lemma, we state the following fact.

**Lemma 4.** *For a CHC system $\mathcal{S}$, let $\mathcal{S}'$ be a system with disequality constraints. Then, if $\langle \mathcal{H}, X_1, \ldots, X_k, Y_1, \ldots, Y_n \rangle \models \mathcal{S}'$, then $\langle \mathcal{H}, X_1, \ldots, X_k, \mathcal{D}_{\sigma_1}, \ldots, \mathcal{D}_{\sigma_n} \rangle \models \mathcal{S}'$ (here $Y_i$ and $\mathcal{D}_{\sigma_i}$ interpret the $diseq_{\sigma_i}$ predicate symbol).*

**Example 3.** For $\mathcal{S} = \{Z \neq S(Z) \to \bot\}$, we get the following system of CHCs:
$$\top \to diseq_{Nat}(Z, S(x))$$
$$\top \to diseq_{Nat}(S(x), Z)$$
$$diseq_{Nat}(x, y) \to diseq_{Nat}(S(x), S(y))$$
$$diseq_{Nat}(Z, S(Z)) \to \bot.$$

Recall that $\mathcal{S}$ is satisfiable in a usual first-order sense, but unsatisfiable in $\mathcal{H}$. But $\mathcal{S}'$ is unsatisfiable in a first-order sense since the query clause is derivable from the first rule, which solves our problem. In our workflow, we *search for finite models of $\mathcal{S}'$ instead of $\mathcal{S}$*, and then act as in the equality-free case. Finally, we end up with the following theorem:

**Theorem 5.** *Let $\mathcal{S}$ be a CHC system and $\mathcal{S}'$ be a CHC system with the disequality constraints. If there is a finite model of $\mathcal{S}'$ over EUF, then there is a regular Herbrand model of $\mathcal{S}$.*

*Proof.* We can rewrite CHCs into DNF, split them into different clauses and eliminate all equality atoms by the unification and substitution. Each clause $C \in \mathcal{S}$ then has the form:
$$C \equiv u_1 \neq t_1 \wedge \ldots \wedge u_k \neq t_k \wedge R_1(\overline{x_1}) \wedge \ldots \wedge R_m(\overline{x_m}) \to H.$$
In $\mathcal{S}'$, this clause becomes $C' \equiv$
$$diseq(u_1, t_1) \wedge \ldots \wedge diseq(u_k, t_k) \wedge R_1(\overline{x_1}) \wedge \ldots \wedge R_m(\overline{x_m}) \to H.$$

So, each clause in $\mathcal{S}'$ has no constraint (rules of $diseq$ have no constraints as well), and by Lemma 2 there is a model $\langle \mathcal{H}, X_1, \ldots, X_k, U_1, \ldots, U_n \rangle$ of every $C' \in \mathcal{S}'$. Then, by Lemma 4 we have $\langle \mathcal{H}, X_1, \ldots, X_k, \mathcal{D}_{\sigma_1}, \ldots, \mathcal{D}_{\sigma_n} \rangle \models C'$. But
$$\langle \mathcal{H}, X_1, \ldots, X_k, \mathcal{D}_{\sigma_1}, \ldots, \mathcal{D}_{\sigma_n} \rangle [\![C']\!] = \langle \mathcal{H}, X_1, \ldots, X_k \rangle [\![C]\!],$$
thus giving us $\langle \mathcal{H}, X_1, \ldots, X_k \rangle \models C$ for every $C \in \mathcal{S}$. □

**On finite model existence for CHCs with the disequality constraints.** There is an interesting observation about finite models and disequality constraints. It can be (straightforwardly) shown that if ADT of sort $\sigma$ has infinitely many terms, then the CHC
$$diseq_\sigma(x, x) \to \bot$$
is satisfied only by infinite structure, i.e., if we force the interpretations of $diseq$ to omit the pairs of equal terms, then such system *has no finite models*. For comparison, if we force $diseq$ to be false in just one tuple, the finite model may exist. For example, the query clause $Q$ over the $Nat$ datatype with
$$Q \equiv diseq_\sigma(Z, Z) \to \bot$$
is satisfiable in a finite model
$$|\mathcal{M}|_{Nat} = \{0, 1\}, \mathcal{M}(Z) = 0, \mathcal{M}(S)(*) = 1,$$
$$\mathcal{M}(diseq_{Nat}) = \{(0, 1), (1, 0), (1, 1)\}.$$
Intuitively, if for proving the satisfiability of CHCs we need to assume the disequality of a large number of ground terms, a finite model is less likely to exist. In practice, this means that tests containing disequality constraints are less likely to be satisfiable in some finite models. This is confirmed by our experimental evaluation in Sec. 8.

### 4.5 Removing Testers and Selectors

In practice, one can get verification conditions not only with constructors of algebraic datatypes, but also with testers and selectors. Unfortunately, a finite-model finder interprets its input as an EUF formula, and thus works in a completely free domain that breaks ADT axioms for testers and selectors. Similarly to the disequality case, we can deal with this problem by preprocessing a set of clauses and replacing testers and selectors with new clauses, e.g., for Lisp-style lists:

$$x = cons(r, y) \rightarrow car(x, r)$$
$$x = cons(y, r) \rightarrow cdr(x, r)$$
$$x = cons(y, z) \rightarrow cons?(x).$$

Thus, a clause

$$\neg(car(x) = cdr(y)) \rightarrow P(x, y)$$

can be rewritten as

$$car(x, a) \wedge cdr(y, b) \wedge \neg(a = b) \rightarrow P(x, y).$$

## 5 Case Study

In this section, we show a verification problem that was solved during experiments with our implementation (see Table 8). This case study demonstrates the expressiveness of regular representations. We also believe that this case may be interesting for readers interested in type theory.

Consider the following program sketch:

```
Var ::= ...
Type ::= arrow(Type, Type)
     | ... <primitive types> ...
Expr ::= var(Var) | abs(Var, Expr)
     | app(Expr, Expr)
Env ::= empty | cons(Var, Type, Env)

fun typeCheck(Γ: Env, e: Expr, t: Type): bool =
    match Γ, e, t with
    | cons(v, t, _), var(v), t -> true
    | cons(_, _, Γ'), var(_), _ ->
        typeCheck(Γ', e, t)
    | _, abs(v, e'), arrow(t, u) ->
        typeCheck(cons(v, t, Γ), e', u)
    | _, app(e₁, e₂), _ ->
        ∃u : Type, typeCheck(Γ, e₂, u) ∧
                   typeCheck(Γ, e₁, arrow(u, t))
    | _ -> false

assert ¬(∃e : Expr, ∀a, b : Type,
    typeCheck(empty, e, arrow(arrow(a, b), a)))
```

This program checks that there is no closed simply typed lambda calculus (STLC) term inhabiting the type $(a \rightarrow b) \rightarrow a$. A quantifier alternation is necessary here to show that there is no term with the *most general*, *principal* [29, def. 3A3] type $(a \rightarrow b) \rightarrow a$. It is well-known that this type is uninhabited, so this program is safe.

We wish to infer an invariant of typeCheck proving the validity of the assertion. Using the weakest liberal precondition calculus [18] we may obtain the verification conditions $VC$ of this program, presented in the Figure 2.

$VC$ is satisfiable modulo theory of algebraic data types Var, Type, Expr and Env, if and only if the program is safe. Moreover, the interpretations of $typeCheck$ satisfying $VC$ are the inductive invariants of the source program.

The strongest inductive invariant of the program is the least fixed point of a step operator, which is the set of all tuples $(\Gamma, e, t)$, such that $\Gamma \vdash e : t$ in STLC typing rules. One needs a very expressive assertion language supporting type theory-specific reasoning to define this invariant. For example, this way is usually used in interactive theorem proving, when the STLC typing is defined in a sufficiently powerful type system of a proof assistant [11].

Instead, our goal is to verify this program automatically, using generic-purpose tools. So it is natural to look for coarser invariants. But does this program have weaker inductive invariants than $\{\langle \Gamma, e, t \rangle \mid \Gamma \vdash e : t\}$, still proving the validity of the assertion[2]? It turns out that the answer is yes, but it is difficult to compose this invariant. One surprisingly simple invariant $\mathcal{I}$ (see below) was discovered by our approach based on the finite model finding engine in CVC4 (see Sec. 4) completely automatically in less than a second.

Every STLC type can be viewed as a propositional formula, where atomic types correspond to atomic variables, and arrows correspond to implications. A *propositional interpretation* $M$ for type $t$ maps atomic variables of $t$ to $\{0, 1\}$. We write $M \models t$ to denote that the propositional interpretation $M$ satisfies the propositional formula corresponding to type $t$. We also say that type $u$ is in $\Gamma \in Env$, if $\Gamma = cons(\ldots, cons(\cdot, u, \ldots)) \ldots)$.

From the Curry-Howard correspondence, we know that the STLC type is inhabited if and only if the propositional formula defined by the type is a tautology of intuitionistic logic. But every intuitionistic tautology is the tautology of classical logic as well. So if type $t$ is inhabited, then $M \models t$ for all propositional interpretations $M$. Thus, the following relation $\mathcal{I}$ over-approximates the strongest inductive invariant of the program:

$$\mathcal{I} \equiv \{\langle \Gamma, e, t \rangle \mid \text{for all } M, \text{ either } M \models t, \text{ or}$$
$$M \not\models u \text{ for some type } u \text{ in } \Gamma\}.$$

Lastly, in our example $(a \rightarrow b) \rightarrow a$ is not a propositional tautology, and $\Gamma$ is empty, so interpreting $typeCheck$ with $\mathcal{I}$ satisfies the last clause of $VC$.

One could attempt to interpret $typeCheck$ with relation

$$\mathcal{J} \equiv \{\langle \Gamma, e, t \rangle \mid t \text{ corresponds to a classical tautology}\},$$

but it fails because $\mathcal{J}$ is not inductive: e.g., it violates the first clause. Conversely, $\mathcal{I}$ satisfies all clauses. We can check that the first clause is satisfied by case splitting: if $M \models t$, then

---

[2]It should be noted that we did not find an answer to this question in the existing literature.

$$\forall \Gamma, \Gamma', e, t, v. \bigl(\Gamma = cons(v, t, \Gamma') \land e = var(v) \to typeCheck(\Gamma, e, t)\bigr) \land$$
$$\forall \Gamma, \Gamma', e, t, t', v, v'. \bigl(\Gamma = cons(v', t', \Gamma') \land e = var(v) \land (v \neq v' \lor t \neq t') \land typeCheck(\Gamma', e, t) \to typeCheck(\Gamma, e, t)\bigr) \land$$
$$\forall \Gamma, e, e', t, t', u, v. \bigl(e = abs(v, e') \land t = arrow(t', u) \land typeCheck(cons(v, t', \Gamma), e', u) \to typeCheck(\Gamma, e, t)\bigr) \land$$
$$\forall \Gamma, e, e_1, e_2, t, u. \bigl(e = app(e_1, e_2) \land typeCheck(\Gamma, e_2, u) \land typeCheck(\Gamma, e_1, arrow(u, t)) \to typeCheck(\Gamma, e, t)\bigr) \land$$
$$\forall e \exists a, b. \bigl(typeCheck(empty, e, arrow(arrow(a, b), a)) \to \bot\bigr)$$

**Figure 2.** Verification conditions $VC$ of the $typeCheck$ program.

$\langle \Gamma, e, t \rangle \in \mathcal{I}$, otherwise $M \not\models t$, but $t$ is in $\Gamma$ by the premise of the clause, so again $\langle \Gamma, e, t \rangle \in \mathcal{I}$. Using the similar dichotomy, it is straightforward to check that $\mathcal{I}$ satisfies the remaining clauses.

Having the inductive invariant $\mathcal{I}$ in hand, it is still challenging to express it in the assertion language. With our new theoretical results for FOL-based languages we formally show in Appendix A that this invariant can not be defined in the first-order theory of algebraic data types Var, Type, Expr and Env, as well as any other safe inductive invariant of this program. For this reason, all state-of-art tools inferring the elementary inductive invariants fail for this program[3].

Instead, we could try to represent $\mathcal{I}$ by a tree automaton. First, there is an automaton, which determines if $t$ is satisfied by a given interpretation $M$. This automaton has two states, 0 and 1. At the $arrow$ constructor, it transits from a pair of states $(1, 0)$ to state 0, and to state 1 from the remaining pairs of states, modeling the logical implication. Starting from states corresponding to the interpretation of the leaves of $t$ by $M$, the automaton stops at state 1 after scanning $t$ iff $M \models t$.

Similarly, we can build an automaton which tests if there is a type $u$ in $\Gamma$, such that $M \not\models u$. It contains two states, $\in$ and $\notin$. At the empty constructor, the automaton transits to the $\notin$ state. At the cons constructor, the automaton transits to the $\in$ state if it is already in the $\in$ state, or it is in the $\notin$ state, and the above automaton stops at 1 for the second argument of cons.

Formally, we have $\{\langle \Gamma, e, t \rangle \mid A \text{ accepts } \langle \Gamma, t \rangle\} \equiv \mathcal{I}$ for the tree automaton $A = \bigl(\{0, 1, \in, \notin, v, e\}, \Sigma_F, \{\langle \in, 0\rangle, \langle \notin, 1\rangle, \langle \in, 1\rangle\}, \Delta\bigr)$ with the following transition relation $\Delta$:

$$\begin{array}{ll}
Var_i \mapsto v & arrow(1, 0) \mapsto 0 \\
PrimType_i \mapsto 0 & arrow(*, *) \mapsto 1 \\
var(v) \mapsto e & empty \mapsto \notin \\
abs(v, e) \mapsto e & cons(v, 1, \notin) \mapsto \notin \\
app(e, e) \mapsto e & cons(v, *, *) \mapsto \in
\end{array}$$

In fact, if we replace the type $(a \to b) \to a$ in the program assertion by an arbitrary type $t$, which is not a tautology of classical logic, $\mathcal{I}$ still would prove the safety of the assertion. We have checked this experimentally. Note that $\mathcal{I}$ is simple enough to completely ignore the type-checked term $e$.

---
[3] Another reason is that the $VC$ contains $\forall \exists$ quantifier alternation in the last clause. Although there are some attempts to design the decision procedure for CHCs with quantifier alternation (e.g., [5]), to the best of our knowledge all of them infer only elementary invariants and cannot handle this example.

One natural question regarding these invariants is what if we try an uninhabited type which corresponds to a classical tautology, but not to an intuitionistic one? One such example is the Pierce's law $t \equiv ((a \to b) \to a) \to a$. In this case $\mathcal{I}$ is too weak to prove that $t$ is uninhabited. Our tool diverged on this input, which might mean that there is no regular inductive invariant, which over-approximates the denotational semantics of typeCheck and still proves the validity of the assertion. In the future, we will investigate it more thoroughly.

## 6 First-Order Invariant Representations

In this section, we compare the expressiveness of the regular representations with the first-order representations of invariants used in the state-of-the-art verification engines. We focus on two invariant representation languages: first-order formulas over ADTs (inferred, for example, by the fixed-point engine in Z3 [36]) and a richer language of first-order formulas over ADTs with size constraints used in Eldarica [32].

Although it is known that tree automata are equivalent to monadic second-order logic over trees [14], which is incomparable to FOL, it is still beneficial to study their relations in detail. Thus, our goal in this section is to come up with a vehicle for disproving relation definability in FOL. Studying the undefinable cases from the practical point of view lets us understand why provers like Z3 or Eldarica diverge for every safe program with undefinable invariants. In this section, we proceed by formulating and proving *pumping lemmas* for two first-order languages.

The concept of pumping lemma arises in the universe of formal languages in connection with finite automata and context-free languages. Pumping lemmas state that for all languages in some class (e.g., regular, context-free), any big enough element can be "pumped", i.e., get an unbounded increase in some of its parts and still stay in the language. Pumping lemmas are useful for proving undefinability: we assume that a language belongs to some class, apply a pumping lemma for that class and get some "pumped" element, which cannot be in the language. This by contradiction proves that the language does belong to that class.

### 6.1 Elementary Representations

We say that a relation $X \subseteq |\mathcal{M}|_{\sigma_1} \times \cdots \times |\mathcal{M}|_{\sigma_n}$ is *first-order definable*, or *elementary*, if there is a formula $\varphi(x_1, \ldots, x_n)$ in the assertion language, such that $(a_1, \ldots, a_n) \in X$ iff $\mathcal{H} \models \varphi(a_1, \ldots, a_n)$. Let Elem be a representation class of elementary relations.

**Example 4** (*IncDec*). Consider the verification conditions of a simple program over Peano integers:
$$x = Z \land y = S(Z) \to inc(x, y)$$
$$x = S(x') \land y = S(y') \land inc(x', y') \to inc(x, y)$$
$$x = S(Z) \land y = Z \to dec(x, y)$$
$$x = S(x') \land y = S(y') \land dec(x', y') \to dec(x, y)$$
$$inc(x, y) \land dec(x, y) \to \bot.$$

The program has an obvious elementary invariant defined by $inc(x, y) \equiv (y = S(x))$, $dec(x, y) \equiv (x = S(y))$. This invariant is the strongest possible, i.e., it expresses the least fixed points of $inc$ and $dec$ correspondingly.

Interestingly, not every program has elementary-definable invariant[4], e.g., the undefinability of the strongest inductive invariant in Example 1 is shown by $E = \{Z, S(S(Z)), \ldots\}$. By extending it with some odd number $E \cup \{S^{2n+1}(Z)\} \subseteq E'$, we violate the query clause with $x = S^{2n}(Z)$ and $y = S^{2n+1}(Z)$. Thus we conclude that $E$ is the only safe inductive invariant.

### 6.2 Pumping Lemma for Elem

In this subsection, we present our results on proving negative elementary definability: a pumping lemma for the first-order languages. We demonstrate it on the *Even* program (Example 1). Then we expand it to the first-order language with size constraints (SizeElem class) and use to prove the undefinability in SizeElem. To the best of our knowledge, this is the first undefinability result for SizeElem.

We begin with auxiliary definitions. The height of a ground term is defined inductively:
$$\mathcal{H}eight(c) \stackrel{\text{def}}{=} 1,$$
$$\mathcal{H}eight(c(t_1, \ldots, t_n)) \stackrel{\text{def}}{=} 1 + \max_{i=1}^{n} \left( \mathcal{H}eight(t_i) \right).$$

For each ADT $\langle \sigma, C \rangle$ and each constructor $f \in C$ having sort $\sigma_1 \times \cdots \times \sigma_n \to \sigma$ for some sorts $\sigma_1, \ldots, \sigma_n$, we define *selectors* $g_i \in S$ with sorts $\sigma \to \sigma_i$ for each $i \leq n$ with the standard semantics: $g_i(f(t_1, \ldots, t_n)) \stackrel{\text{def}}{=} t_i$.

A *path* is a (possibly empty) sequence of selectors: for $S_n : \sigma_n \to \sigma_{n-1}, \ldots, S_1 : \sigma_1 \to \sigma_0$, we define a path $s \stackrel{\text{def}}{=} S_1 \ldots S_n$. For terms $t$ of sort $\sigma_n$, we define $s(t) \stackrel{\text{def}}{=} S_1(\ldots(S_n(t))\ldots)$. For ground terms $g$, we redefine $s(g)$ to be a computed subterm of $g$ at $s$. We denote paths with small letters $p, q, r, s$.

We say that two paths $p$ and $q$ *overlap* if one of them is the suffix of the other. For pairwise non-overlapping paths $p_1, \ldots, p_n$, by $t[p_1 \leftarrow u_1, \ldots, p_n \leftarrow u_n]$ we denote the term obtained by simultaneous replacement of subterms $p_i(t)$ by $u_i$ in $t$. For finite sequence of pairwise-distinct paths $P = (p_1, \ldots, p_n)$ and some terms $U = (u_1, \ldots, u_n)$, we abuse the notation and write $t[P \leftarrow U]$ meaning $t[p_1 \leftarrow u_1, \ldots, p_n \leftarrow u_n]$ and also $t[P \leftarrow t]$ for $t[p_1 \leftarrow t, \ldots, p_n \leftarrow t]$.

[4]In general, a program might have more than one inductive invariant. However, examples in this paper are designed to have unique inductive invariant.

Now we define a set of paths, which would be pumped.

**Definition 4.** For all sorts $\sigma$, we say that a term $t$ is a *leaf term* of sort $\sigma$ if it is either a base constructor, or $t = c(t_1, \ldots, t_n)$, where all $t_i$ are leaf terms and $t$ does not contain any proper subterms of sort $\sigma$. For a ground term $g$ and a sort $\sigma$, we define $leaves_\sigma(g) \stackrel{\text{def}}{=} \{p \mid p(g) \text{ is a leaf term of sort } \sigma\}$.

**Lemma 6** (Pumping Lemma for Elem). *Let L be an elementary language of n-tuples. Then, there exists a constant $K > 0$ satisfying: for every n-tuples of ground terms $\langle g_1, \ldots, g_n \rangle \in$ L, for any $i$ such that $\mathcal{H}eight(g_i) > K$, for all infinite sorts $\sigma \in \Sigma_S$ and for all paths $p$ with a length greater than $K$, there exist finite sets of paths $P_j$ such that $p \in P_i$, for all $p_1, p_2 \in \bigcup_j P_j$ it is true that $p_1(g) = p_2(g)$, and there is $N \geq 0$, such that for all $t$ of sort $\sigma$ with $\mathcal{H}eight(t) > N$ it holds that:*
$$\langle g_1[P_1 \leftarrow t], \ldots, g_i[P_i \leftarrow t], \ldots, g_n[P_n \leftarrow t] \rangle \in \mathbf{L}.$$

*Proof.* See Appendix B.1. □

Intuitively, Lemma 6 says that for big enough tuples of terms, we can take any of the deepest subterms, substitute them with *arbitrary* term $t$, and *still* obtain a tuple of terms in the language. The lemma formalizes our intuition that the first-order language of ADTs can only describe equalities and disequalities between subterms of a bounded depth: if one goes deep enough and replaces the leaf terms with the arbitrary terms, the initial and the resulting terms will be *indistinguishable* by the first-order language.

Now we demonstrate how to use the lemma to prove the undefinability results. Using pumping lemma, we prove that the invariant in the *Even* example is non-elementary.

**Proposition 1.** *Even* $\notin$ Elem.

*Proof.* Assume that there is the elementary safe inductive invariant L. Take a constant $K > 0$ from the pumping lemma. Let $g \equiv S^{2K}(Z) \in \mathbf{L}, \sigma = Nat, p = S^{2K}$. Now, $\bigcup_j leaves_\sigma(g_j) = leaves_\sigma(g) = \{p\}$, so $P = \{p\}$. Then, by pumping lemma there is $N \geq 0$, and we take $t \equiv S^{2N+1}(Z)$. By the lemma $g[P \leftarrow t] \equiv S^{2K}(S^{2N+1}(Z)) \in \mathbf{L}$. But then L violates the last clause of the verification conditions, thus L is not a safe inductive invariant, contradiction. □

As we have shown, pure elementary representations cannot express some program invariants. There are some attempts to increase the expressiveness of the first-order language by extending it with some additional symbols, still keeping the satisfiability of its formulas decidable.

### 6.3 Elementary Representations with Size Constraints

One natural idea to increase the expressiveness of the elementary representations is to extend the first-order language with the ability to specify *term sizes*. The inference of inductive invariants representable in that class is automated in the Eldarica CHC solver [32].

The language of SizeElem class can be derived from the language of Elem by adding into the signature a sort $Int$, Presburger arithmetic operations and function symbols $size_\sigma$ with arity $\sigma \to Int$, counting constructors in a term of sort $\sigma$. We omit $\sigma$ from $size$ symbols for brevity.

The satisfiability of formulas with size constraints is checked in the $\mathcal{H}_{size}$ structure, obtained by conjoining the standard model of Presburger arithmetic to $\mathcal{H}$ and interpreting $size_\sigma$ function symbols straightforwardly. For example, given term
$$t \equiv cons(Z, cons(S(Z), nil))$$
with the sort $NatList := nil : NatList \mid cons : Nat \times NatList \to NatList$, we have $\mathcal{H}_{size}[\![size(t)]\!] = 6$.

We incorporate notations of Hojjat and Rümmer [31] and denote $\mathbb{T}_\sigma^k = \{t \text{ has sort } \sigma \mid size(t) = k\}$. For each ADT sort $\sigma$ we define the set of term sizes $\mathbb{S}_\sigma = \{size(t) \mid t \in |\mathcal{H}|_\sigma\}$. A *linear set* is a set of the form $\{\mathbf{v} + \sum_{i=1}^n k_i \mathbf{v}_i \mid k_i \in \mathbb{N}_0\}$, where all $\mathbf{v}, \mathbf{v}_i$ are vectors over $\mathbb{N}_0 = \mathbb{N} \cup \{0\}$.

**Definition 5.** An ADT sort $\sigma$ is *expanding* iff for every natural number $n$ there is a bound $b(\sigma, n) \geq 0$ such that for each $b' \geq b(\sigma, n)$, if $\mathbb{T}_\sigma^{b'} \neq \varnothing$, then $\left|\mathbb{T}_\sigma^{b'}\right| \geq n$. An ADT signature is called expanding if all its sorts are expanding.

**Example 5** (*EvenLeft*). Consider a tree datatype $Tree := leaf : Tree \mid node(Left : Tree, Right : Tree)$ and the program *EvenLeft* which checks if the leftmost branch of the tree has even number of nodes:

$$x = leaf \to EvenLeft(x)$$
$$x = node(node(x', y), z) \land EvenLeft(x') \to EvenLeft(x)$$
$$EvenLeft(x) \land EvenLeft(node(x, y)) \to \bot$$

Noteworthy, SizeElem admits quantifier elimination [58], so intuitively size constraints can only make restrictions on some subterms of finite depth, and they count all constructors at a time — we cannot specify to count only "the leftmost branch" constructors, as in example. More formally:

**Lemma 7** (Pumping Lemma for SizeElem). *Let the ADT signature be expanding and let* $\mathbf{L}$ *be an elementary language of $n$-tuples with size constraints. Then, there exists a constant $K > 0$ satisfying: for every $n$-tuple of ground terms $\langle g_1, \ldots, g_n \rangle \in \mathbf{L}$, for any $i$, such that $Height(g_i) > K$, for all infinite sorts $\sigma \in \Sigma_S$, and for all paths $p \in leaves_\sigma(g_i)$ with length greater than $K$, there exists an infinite linear set $T \subseteq \mathbb{S}_\sigma$, such that for all terms $t$ of sort $\sigma$ with sizes $size(t) \in T$, there exist sequences of paths $P_j$, with no path in them being a suffix of path $p$, and sequences of terms $U_j$, such that*

$$\langle g_1[P_1 \leftarrow U_1], \ldots, g_i[p \leftarrow t, P_i \leftarrow U_i], \ldots, g_n[P_n \leftarrow U_n] \rangle \in \mathbf{L}.$$

*Proof.* See Appendix B.2. □

Intuitively, having a SizeElem language, a big enough term $g$ from it and deep enough path $p$, one can substitute $p(g)$ with almost arbitrary term $t$, limited only by size being in some linear, but infinite set $T$, and still obtain the term of the language. This means that for SizeElem languages there are unconstrained subterms, which are thus indistinguishable in this representation. Now we demonstrate the pumping lemma in action. To the best of our knowledge, this is the first negative definability result for SizeElem.

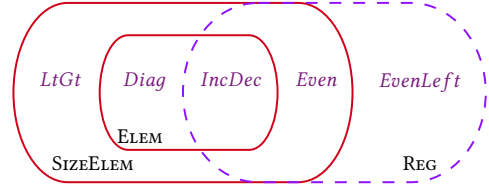

**Figure 3.** The comparison of the expressive power of three computable representations of inductive invariants. *LtGt* and *Diag* verification conditions can be found in the Appendix C.

**Proposition 2.** *EvenLeft* $\notin$ SizeElem.

*Proof.* First, $Tree$ sort is expanding. Let us assume *EvenLeft* is in SizeElem and it has an invariant $\mathbf{L}$. Take $K > 0$ from the lemma. Let $g \in \mathbf{L}$ be a full binary tree of height $2K$, $\sigma = Tree, p = Left^{2K}$. Take the infinite set of positive integers $T$ from the lemma. We can take some $n \in T, n > 2$ and $t = node(leaf, t')$ for some $t'$, such that $size(t) = n$. Now by the lemma, there exists a sequence of paths $P$ and a sequence of terms $U$, and none of elements in $P$ is a suffix of $p$. By the lemma, we must have $g[p \leftarrow t, P \leftarrow U] \in \mathbf{L}$, so the leftmost tree path must have an even length. However, at the leftmost path $p = Left^{2K}$ there is a term $node(leaf, t')$, so the leftmost tree path has length $2K - 1 + 2 = 2K + 1$, which is odd. Contradiction. □

In theory, size constraints add the expressive power to the assertion language. But in practice, it might happen that the automated inference of SizeElem invariants becomes harder because of checking the validity and generalization of size constraints. We evaluate Eldarica against other automated solvers to check that in Sec. 8.

## 7 Comparison of the Expressive Power

In this section, we compare the expressiveness of Elem, Reg and SizeElem classes. Our results are summarized in Figure 3. The regions in figure denote the representation classes: the small red one corresponds to the Elem class, the big red one stands for the SizeElem class, and the dotted purple region denotes the Reg class. The names in italic font denote programs. Each program inside a region denotes that the program has a safe inductive invariant definable in the corresponding class.

Due to the space limits, we skip the detailed proofs of the results in this picture. Instead, they can be found in Appendix C. Two proofs using our pumping lemmas have already been presented in Sec. 6.

Our results show that Reg and Elem classes, as well as Reg and SizeElem classes, are incomparable: there are programs with invariants definable in both classes and programs with invariants definable only in one of these classes. A quick takeaway message is that first-order, logic-based representations and automata-based representations of relations in Herbrand universe are very different. While automata may specify precise enough properties of terms on unbounded depth, they fail to represent *relational properties* on terms (like disequality of terms or orderings over Peano numbers). On the other hand, while elementary representations are good for expressing the equality and disequality relations of terms, but can be precise only on a bounded depth.

**Future Work.** Regular relations can be further extended to other tree language classes (see, e.g., [9, 10, 20, 25, 33, 39]). Another intriguing prospect is to automate the inference of inductive invariants represented in first-order languages with regular language membership predicates [15]. This language is known to be decidable and closed under Boolean operations and subsumes both Reg and Elem classes.

## 8 Implementation and Experiments

We have evaluated our tool inferring regular invariants against state-of-the-art: Z3 inferring the Elem invariants and Eldarica using SizeElem representations on existing benchmarks. The amount of solved tasks correlates with definability: if a program has no invariants definable in a class, the corresponding solver diverges.

**Implementation.** We have implemented a regular invariant inference tool called RInGen[5] based on the preprocessing approach presented in Sec. 4. RInGen accepts input clauses in the SMTLIB2 [3] format and TIP extension with `define-fun-rec` construction [12]. It takes conditions with a property and checks if the property holds, returning SAT and the safe inductive invariant if it does or terminates with UNSAT if it does not. Thus RInGen can be run as a backend solver for functional program verifiers, such as MoCHi [35] and RCaml [55]. We run CVC4[6] as a backend multi sort finite-model finder to find regular models (Reg in our notation, see Example 3). Besides regular models, a finite model finding approach of CVC4 [50] based on quantifier instantiation provides us with sound unsatisfiability checking.

**Benchmarks.** We have empirically evaluated RInGen against state-of-the-art CHC solvers on "Tons of inductive problems" (*TIP*) benchmark set by Claessen et al. [12] and our own benchmarks inspired by the benchmark of De Angelis et al. [16]. The *TIP* benchmark set was preprocessed to be compatible with CHC solvers and the modified version[7] was contributed to the CHC-COMP solver competition[8].

We have modified the benchmark of De Angelis et al. [16] by replacing all non-ADT sorts with ADTs and adding CHC-definitions for non-ADT operations. The aggregated test set[9] consists of 60 CHC systems over binary trees, queues, lists, and Peano numbers. The test set was divided into two problem subsets *PositiveEq* and *Diseq*. *PositiveEq* is a set of CHC-systems with equality occurring only positively in clause bodies. *Diseq* set includes tests with occurrences of disequality constraints in clause bodies, substituted with `diseq` atoms, which is a sound transformation (see Sec. 4.4).

---

[5]*R*egular *In*variant *Gen*erator: https://github.com/Columpio/RInGen.
[6]Using `cvc4 --finite-model-find`.
[7]https://github.com/chc-comp/ringen-adt-benchmarks.
[8]https://chc-comp.github.io/.
[9]https://gitlab.com/Columpio/adt-benchmarks/.

The *TIP* benchmark consists of 454 inductive ADT problems over lists, queues, regular expressions, and Peano integers originally generated from functional programs. From the original benchmark [12], we filtered out problems with only ADT sorts (the remaining problems use the combinations of ADTs with other theories), converted all of them into CHCs, replaced the disequalities with the `diseq` atoms as described in Sec. 4.4 and replaced all free sorts declared via (`declare-sort` ... 0) with the *Nat* datatype.

**Competing tools.** The evaluation was performed against Z3/Spacer [17] with the Spacer engine [37] and Eldarica [32] — state-of-the-art Horn-solvers which construct elementary models and support ADTs. Spacer works with elementary model representations (Elem in our notation, see Sec. 6.1). It incorporates standard decision, interpolation and quantifier elimination techniques for ADT [6]. Spacer is based on *property-directed reachability* (PDR), which alternates subtasks of counter-example finding and safe invariant construction by propagating reachability facts and pushing partial safety lemmas in a property-directed way.

Eldarica builds models with size constraints, which count the total number of constructor occurrences in them (named SizeElem in our notation, recall Sec. 6.3). It relies on the Princess SMT solver [51], which offers decision and interpolation procedures for ADT with size constraints by reduction to the combination of EUF and LIA [31].

As a baseline we include the CVC4 induction solver [49] into the comparison (denoted CVC4-Ind[10]), which leverages a number of techniques for inductive reasoning in SMT.

Finally, we provide a comparison against a VeriMAP extension called VeriMAP-iddt [16]. It handles the verification conditions over LIA and ADT and eliminates ADTs from the verification conditions completely by applying the fold/unfold techniques, so it is a clause transformer, which does not produce invariants over ADTs. Thus, it does not serve our main goal of comparing the expressivity of different invariant classes for ADT, however we include it as a baseline.

Experiments were performed on an Arch Linux machine Intel(R) Core(TM) i7-4710HQ CPU @ 2.50GHz 2.50GHz processor with 16GB RAM and a 300-second timeout.

**Results.** The results are summarized in Table 1.

On *PositiveEq* and *Diseq* benchmark sets, Spacer solved 7 problems and for the rest it ended with 8 UNKNOWN results and 45 timeouts. Eldarica solved 2 problems (also solved by Spacer) with 58 timeouts. RInGen found 31 regular solutions, one counterexample and had 28 timeouts. Most of the solved problems are from *PositiveEq* test set, which does not contain equalities in the negative context. This confirm our hypothesis that such problems more likely have regular invariants, which is discussed in Sec. 4.4. Each problem solved by Spacer or Eldarica was solved by RInGen as well.

---

[10]Using `cvc4 --quant-ind --quant-cf --conjecture-gen --conjecture-gen-per-round=3 --full-saturate-quant`

**Table 1.** Results of experiments on three ADT problem sets. Number in each cell stands for the amount of correct results within 300-seconds time limit. RInGen was used for *regular model* construction, Spacer was used for *elementary model* construction and Eldarica was used for building *elementary models with size constraints*. CVC4-Ind and VeriMAP-iddt are included for reference.

| Invariant Representation | | | Reg | SizeElem | Elem | - | - |
|---|---|---|---|---|---|---|---|
| Problem Set | # | Answer | RInGen | Eldarica | Spacer | CVC4-Ind | VeriMAP-iddt |
| *PositiveEq* | 35 | SAT | 27 | 1 | 4 | 0 | 0 |
| *Diseq* | 25 | SAT | 4 | 0 | 2 | 0 | 2 |
| | | UNSAT | 1 | 1 | 1 | 1 | 1 |
| *TIP* | 454 | SAT | 30 | 46 | 26 | 0 | 16 |
| | | Unique SAT | 13 | 25 | 7 | 0 | 0 |
| | | UNSAT | 21 | 12 | 22 | 13 | 10 |
| | | Unique UNSAT | 5 | 0 | 5 | 0 | 0 |
| Total | 514 | SAT | 61 | 47 | 32 | 0 | 18 |
| | | UNSAT | 22 | 13 | 23 | 14 | 11 |

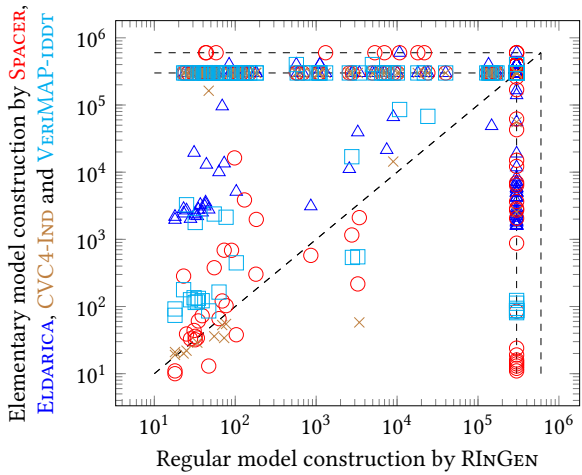

**Figure 4.** Comparison of engines performance. Each point in a plot represents a pair of the run times (sec × sec) of RInGen for Reg construction (x-axis) and a competitor for (Size)Elem construction (y-axis). Timeouts are placed on the inner dashed lines, crashes are on the outer dashed lines.

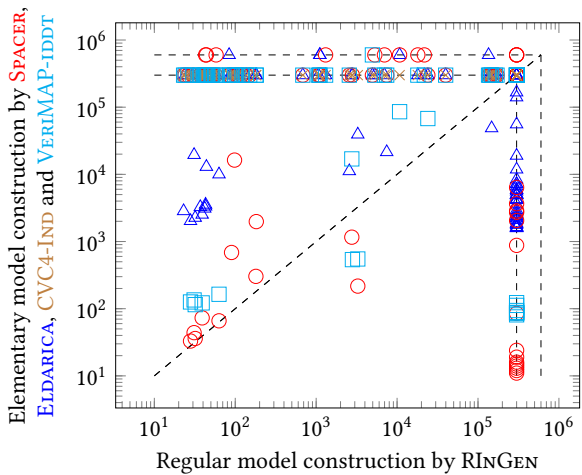

**Figure 5.** Comparison of engine performance with *only SAT results shown*. The testcase is included into this plot, if at least one of engines has discovered an invariant.

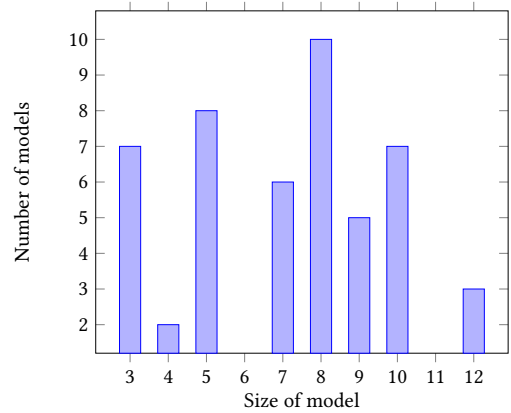

**Figure 6.** Sizes of finite models found in the evaluation. The size of model (x-axis) is calculated as the *sum* of all sorts cardinalities.

The *TIP* benchmark gave more diverse results. Firstly, all 12 problems claimed to be UNSAT by Eldarica were covered by RInGen as well, i.e., RInGen managed to find counterexamples more efficiently than Eldarica. RInGen and Spacer witnessed the unsatisfiability of 21 and 22 CHC-systems respectively. Most of these problems intersect, although some of them are unique for each solver.

Spacer exceeded the time limit 393 times. It terminated 13 times within time limit with the UNKNOWN result. RInGen exceeded the time limit 403 times, and Eldarica stopped after the time limit 368 times with 28 errors[11].

Finally, Eldarica proved safety with SAT result in 46 cases vs 30 cases of RInGen. They share 11 problems, on which RInGen was two magnitudes faster. RInGen has 13 uniquely solved problems, all of them with some variant of evenness predicate on Peano numbers (e.g., "the length of list concatenation is even iff sum of list lengths is even"). This type of regularity is naturally handled by the finite-model finder (see Prop. 7 and Prop. 10). Eldarica has 25 uniquely solved problems, all of them with orderings ($<, \leq, >, \geq$) on Peano numbers. Note that this is exactly the case considered in Prop. 12, differing SizeElem from the rest classes.

---
[11]All with the same message: `"Cannot handle general quantifiers in predicates at the moment"`.

Timing plots in Figure 4 and Figure 5 show that not only RInGen inferred more invariants but it also was generally faster than other tools. On Figure 4, some unsafe benchmarks were handled faster by CVC4-Ind, VeriMAP-iddt and Spacer. This is possibly due to a more effective procedure of quantifier instantiation in CVC4-Ind and a more balanced tradeoff between the invariant inference and the counterexample search in the PDR core of Spacer (which is also called by the VeriMAP-iddt).

A diagram in Figure 6 shows the model sizes produced by the CVC4-f finite-model finder during the experiments.

**Other experiments.** We have tried 23 hand-written programs related to the type theory (recall Sec. 5), questioning the inhabitance of different STLC types, typability of STLC terms, and programs modeling different term-rewriting systems. All these benchmarks were intractable for all the solvers, except the finite model finder. For that reason, we omit the detailed statistics. We have also tried to run another finite model finders (for example, Mace4) as a backend, but they have shown worse results than CVC4.

**Discussion.** Clearly, finite model finding did much better on our own benchmarks inspired by the one of De Angelis et al. [16]. This is due to two reasons: the expressiveness of tree automata for representing the invariants and the efficiency of RInGen's backend CVC4-f engine. More importantly, Spacer and Eldarica diverged more often because of inexpressiveness of their FOL-based languages. Within the limits of their invariant representations, they perform smoothly.

On *TIP* benchmarks, Eldarica solved more testcases than RInGen, but the analysis of the testcases solved only by Eldarica has shown, that all such tests define the Peano ordering, easily handled by Eldarica by the reduction to LIA. On testcases solved by both engines, RInGen was faster in average. Still, the majority of interesting test cases in the *TIP* set obtained from proof assistants is currently beyond the reach of state-of-the-art engines under comparison, which motivates our future work.

To sum up, tree automata have been proven to be a promising technique in automated verification of ADT-manipulating programs. They allow to express complex properties of the recursive computation and can be efficiently inferred by the existing engines. In the future, however, a hybrid approach to infer invariants in parts by automata and in parts by FOL should exhibit the best performance.

## 9 Related Work

Language classes considered in this work have already been studied in the literature. Although these were separate works from different subfields of computer science.

**Finite models and tree automata.** A classic book on automated model building [8] gives a generous overview of finitely representable models and their features, decision procedures and closure properties. In the context of invariant inference it is important that checking the inductiveness of a candidate finite-model invariant is decidable and checking the existence of a finite model is not. The latter is semidecidable: we eventually terminate if a finite model exists, but do not terminate if the system has only infinite models.

Basic results for tree automata are accumulated in [14]. There is also an ongoing research on extensions of regular tree languages, which still enjoy nice decidability and closure properties [9, 10, 20, 25, 33, 39].

A number of tools like Mace4 [44], Finder [52], Paradox [13] and CVC4 [50] are used to find finite models of first-order formulas. Most of them implement a classic DPLL-like search with propagating assignments. CVC4, in addition, uses conflict analysis to accelerate the search. They were applied to various verification tasks [40] and even infinite models construction [46]. Yet we are unaware of applying finite model finders to inference of invariants of ADT-manipulating programs.

Recently, Haudebourg et al. [28] proposed a regular abstract interpretation framework for invariant generation for higher-order functional programs over ADTs. Authors derive a type system where each type is a regular language and use CEGAR to infer regular invariants. Their procedure is much more complex because they support high-order reasoning which is not the goal of this paper, comparing ADT-invariant representation. Targeting first-order functions over ADT only we obtain a more straightforward invariant inference procedure by using effective finite-model finders. Moreover, our work clarifies the gap between different invariant representations and their expressivity and aims not to advertise regular invariants themselves but to overcome mental inertia towards elementary invariant representations.

**Herbrand model representations.** There is a line of work studying different computable representations of Herbrand models [22, 23, 26, 54], which can be fruitful to study to find out new ADT invariant representations. Even though tree automata enjoy lots of effective properties, they have limited expressivity, so their extensions were widely studied in the automated model building field [8]. A survey on computational representations of Herbrand models, their properties, expressive power, correspondences and decision procedures can be found in [42, 43].

**ADT solving.** There is a plenty of proposed quantifier elimination algorithms and decision procedures for the first-order ADT fragment [4, 45, 47, 48, 53] and for an extension of ADT with constraints on term sizes [58]. Some works discuss the Craig interpolation of ADT constraints [31, 34]. Such techniques are being incorporated by various SMT solvers, like Z3 [17], CVC4 [2], and Princess [51].

Some work on automated induction for ADT was proposed. Support for inductive proofs exists in deductive verifiers, such as Dafny [38] and SMT solvers [49]. The technique in CVC4 is deeply integrated in the SMT level — it implements Skolemization with inductive strengthening and term enumeration to find adequate subgoals. De Angelis et al. [16] introduces a fold/unfold-based technique for eliminating ADTs from the CHC-system by transforming it to CHC-system over LIA and booleans. Recently, Yang et al. [57] applied a method based on Syntax-Guided Synthesis [1] to leverage induction by generating supporting lemmas based on failed proof subgoals and user-specified templates. On the whole, checking the inductiveness of FOL invariants is decidable as FOL is a decidable theory. Invariant inference is semidecidable in the sense that we can enumerate all candidate FOL-formulas but we do not terminate if the system has no FOL-representable invariant.

**ADTs with size constraints.** A brief survey of ADT theory with sized constraints is proposed by Zhang et al. [58]. In particular, quantifier elimination and decision procedures for this class are introduced. The automated inference of inductive invariants in SizeElem is implemented in Eldarica Horn solver [32]. The ADT constraints are solved and generalized by the Princess SMT solver [51], using the reduction to uninterpreted functions and linear integer arithmetic [31] with handling size constraints in spirit of Suter et al. [53].

## 10  Conclusion

We have compared the two branches of representing the invariants of programs manipulating ADTs: by the first-order formulas over ADTs, possibly extended with size constraints, or by tree automata. We have shown that these branches are incomparable, each with its own downsides and upsides.

We have demonstrated that tree automata are very promising for representing the invariants of computation over ADTs, as they allow to express properties of the unbound depth. On the downside, tree automata cannot express the relations between different variables.

Using the correspondence between finite models and tree automata, we were able to use the existing finite model finders for automated inference of regular inductive invariants. We have bypassed the problem of disequality constraints in the verification conditions and implemented a tool called RInGen which automatically infers the regular invariants of ADT-manipulating programs. This tool is competitive with the state-of-art CHC solvers Z3/Spacer and Eldarica. Using RInGen, we have managed to detect interesting invariants of various inductive problems, including the non-trivial invariant of the inhabitance checking for STLC.

In the future, we mainly plan to investigate extensions of tree automata which subsume elementary invariants but still enjoy efficient decidability and closure properties.

## A  Analysis of the Case Study

In this appendix, we demonstrate the effectiveness of the "pumping" lemma machinery, introduced in Sec. 6.2, in proving negative definability for FOL-based languages. The demonstration is performed on a sophisticated $STLC$ type inhabitation example from Sec. 5.

For readability, let us denote $STLC$ terms and types in the classical manner: $\lambda x.\lambda y.x$, $a \rightarrow a$ and so on.

**Proposition 3.** *The CHC-system STLC from Figure 2 does not have a FOL-definable invariant.*

*Proof.* This fact can be proven by the application of the pumping lemma for the Elem class. Let us prove that by contradiction. Suppose there is a FOL-definable invariant called $\mathbf{L}$. With a constant $K$ from the pumping lemma we can define
$$e \stackrel{\text{def}}{=} \lambda i_1.\ldots.\lambda i_{K+1}.\lambda x.\lambda y.x.$$
Then for arbitrary ground term $a$ of the ADT-sort $Type$ we can define
$$t_1 \stackrel{\text{def}}{=} \underbrace{(a \rightarrow a) \rightarrow \ldots \rightarrow (a \rightarrow a)}_{K+1 \text{ times}} \rightarrow (a \rightarrow a \rightarrow a).$$
It is easy to see from the typing rules that $\langle empty, e, t_1 \rangle \in \mathbf{L}$. Also $\mathcal{H}eight(t_1) > K$.

Let $L$ and $R$ also denote a domain and a codomain selectors of the typing arrow. Let us take $\sigma = Type$ and $p = R^{K+1}$, so $p(t_1) = (a \rightarrow a \rightarrow a)$. The lemma states that there should be finite sets of paths $P_j$ such that $p \in P_i$ for some $i$, for all $p_1, p_2 \in \bigcup_j P_j$ it is true that $p_1(t_1) = p_2(t_1)$. Let us denote $P \stackrel{\text{def}}{=} \bigcup_j P_j$. As $p(t_1) = (a \rightarrow a \rightarrow a)$ and there is no such subterm elsewhere in $t_1$, $P = \{p\}$. For $N \geq 0$ from the lemma then we can define
$$t_2 \stackrel{\text{def}}{=} \underbrace{(a \rightarrow a) \rightarrow \ldots \rightarrow (a \rightarrow a)}_{N \text{ times}} \rightarrow a.$$
The lemma then states that $\langle empty, e, t_1[P \leftarrow t_2] \rangle \in \mathbf{L}$. Consider that
$$t_1[P \leftarrow t_2] = t_1[p \leftarrow t_2] = \underbrace{(a \rightarrow a) \rightarrow \ldots \rightarrow (a \rightarrow a)}_{K+N+1 \text{ times}} \rightarrow a.$$
By applying STLC application rule $K + N + 1$ times to $e$ and $\lambda x.x$ term we obtain
$$\langle empty, e \underbrace{(\lambda x.x) \ldots (\lambda x.x)}_{K+N+1 \text{ times}}, a \rangle \in \mathbf{L}.$$
Thus, there exists a term which for any type $a$ is typed with $a$ in the empty context by $\mathbf{L}$. In particular, this term is also typable by $\mathbf{L}$ with $(a \rightarrow b) \rightarrow a$ for arbitrary $a$ and $b$, which contradicts the goal clause of the $STLC$ CHC system. $\square$

## B  Proofs of Lemmas

We begin with definition of an extended first-order ADT language with testers and selectors. As was shown by Oppen [45], such extended language admits quantifier elimination.

An (extended) algebraic data type is a tuple $\langle C, S, T, \sigma \rangle$, where $\sigma$ is a sort, $C$ and $S$ are sets of uninterpreted function symbols (called constructors and selectors, respectively), $T$ is a set of predicate symbols (called testers), and $C, S, T$ do not have common elements. For every $c \in C$, if $c$ has sort $\sigma_1 \times \cdots \times \sigma_n \rightarrow \sigma$, then for all $i = 1...n$ there are selector function symbols $g_i \in S$ with sorts $\sigma \rightarrow \sigma_i$ with standard semantics
$$g_i(f(t_1, \ldots, t_n)) \stackrel{\text{def}}{=} t_i.$$
As all selectors are unary, we write $S_1 \ldots S_n x$ instead of $S_1(\ldots(S_n(x)))$ for selectors $S_i$ to avoid unnecessary parenthesis. We also define $\|s\| \stackrel{\text{def}}{=} n$ for paths $s \stackrel{\text{def}}{=} S_1 \ldots S_n$. For each $c \in C$ there is a testing predicate symbol $c? \in T$ with sort $\sigma$ with predefined interpretation
$$c?(t) \Leftrightarrow \exists t_1, \ldots, t_n.t = c(t_1, \ldots, t_n).$$
We assume that $\sigma_i \neq \sigma_j$ and for $i \neq j$ and $C_i, T_j, S_k$ are pairwise disjoint and put the signature $\Sigma = \langle \Sigma_S, \Sigma_F, \Sigma_P \rangle$, where $\Sigma_S = \{\sigma_1, \ldots, \sigma_n\}$, $\Sigma_F = C_1 \cup \cdots \cup C_n \cup S_1 \cup \cdots \cup S_n$ and $\Sigma_P = \{=_{\sigma_1}, \ldots, =_{\sigma_n}\} \cup T_1 \cup \cdots \cup T_n$.

### B.1  Pumping Lemma for Elem

**Definition 6** (Normal Form for Elem). *A $\Sigma$-formula $\varphi(x_1, \ldots, x_n)$ is in a normal form iff it is in DNF, quantifier-free and each of it's atoms has one of the following forms:*

- *$c?(s(x))$, for some tester $c?$, path $s$ and variable $x$,*
- *$s(x) = s'(y)$, for some paths $s, s'$ and variables $x, y$,*
- *$s(x) = c$, for some path $s$, a leaf constructor $c$ and variable $x$.*

*Each conjunct in DNF should also be not equivalent to $\bot$ and for each subterm $S(t)$, where $S$ is a selector for a constructor $c$, there should be an atom $c?(t)$.*

It easy to see that each formula has an equivalent formula in a normal form, as the theory of ADT admits quantifier elimination and all constructors can be eliminated by introducing selectors. For example, for LISP-style lists:
$$cons(a, b) = y \Leftrightarrow cons?(y) \wedge car(y) = a \wedge cdr(y) = b,$$
$$car(cons(a, b)) = y \Leftrightarrow a = y.$$

We call a language of $n$-tuples $n$-dimensional. We can prove the pumping lemma for Elem for general case of $n$-dimension languages by proving 1-dimensional case which can be reformulated as follows.

**Lemma 8** (Pumping Lemma for Elem (one-dimensional)). *Let $\mathbf{L}$ be a 1-dimensional elementary language. Then, there exists a constant $K > 0$ satisfying: for every ground $g \in \mathbf{L}$ with $\mathcal{H}eight(g) > K$, for all infinite sorts $\sigma \in \Sigma_S$ and for all $p$ such that $\|p\| > K$, there exists a finite sequence of paths $P$ such that $p \in P$, for all $p_1, p_2 \in P$ it is true that $p_1(g) = p_2(g)$, and there is $N \geq 0$, so that for all $t$ of sort $\sigma$ with $\mathcal{H}eight(t) > N$ holds $g[P \leftarrow t] \in \mathbf{L}$.*

*Proof.* By definition, for **L** there is some formula $\varphi(x)$, defining the language. Without loss of generality, let us assume that it is in a normal form and $\varphi(x) \equiv \bigvee_i \varphi_i(x)$.

Let the constant $K$ be the sum of size of the formula $\|\varphi\|$ (defined syntactically, as number of all symbols, logical and non-logical) and the maximum size of all leaf terms in Herbrand universe (obviously, it is finite). Now let $g \in \mathbf{L}$ be some term such that $\mathcal{H}eight(g) > K$, let $\sigma \in \Sigma_S$ be an infinite sort, and path $p \in leaves_\sigma(g)$ be a path with $\|p\| > K$.

As $g \in \mathbf{L}$, there is some $\varphi_i$ which is satisfied on $g$. We will construct $P$, so that $g[P \leftarrow t]$ is still satisfied on $\varphi_i$ and thus on $\varphi$, proving the lemma. First, as $\varphi$ is in a normal form, $\varphi_i(x)$ can be represented as

$\varphi_i(x) \equiv \Phi_=(x) \wedge \Phi_{\neq}(x) \wedge \Phi_{=c}(x) \wedge \Phi_{\neq c}(x) \wedge \Phi_{c?}(x) \wedge \Phi_{\neg c?}(x)$,

where $\Phi_{=c}(x)$ and $\Phi_{\neq c}(x)$ contain only positive and negative literals with equalities with constant constructor on one side correspondingly, $\Phi_=(x)$ and $\Phi_{\neq}(x)$ contain only positive and negative literals with equalities of the form $s(x) = s'(x)$, $\Phi_{c?}(x)$ and $\Phi_{\neg c?}(x)$ contain only positive and negative testers.

From all equalities in $\Phi_=(x)$ we can construct a bidirectional congruence closure graph over *paths* in the manner of Oppen [45]. For each path $q$ from the graph which is a suffix of $p$, let $r_q$ be a path, s.t. $p = r_q q$, and $E_q$ to be a set of all paths which are equal to $q$, according to the graph. Then, we put

$P \stackrel{\text{def}}{=} \{r_q e \mid q \text{ is a suffix of } p \text{ from the graph}, e \in E_q\}$.

For example, consider ADT $Tree = node(Tree, Tree) \mid leaf$ with selectors $L$ and $R$ for left and right subterms of the *node* correspondingly. Then, for a formula $\Phi_=(x) \equiv LLx = RRx \wedge LRx = RRx$ and a path $p = RRLR$, we have a bidirectional closure graph with three vertices:

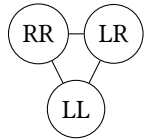

So there is only one $q = LR$ from the graph, which is the suffix of the $p = RRLR = r_q q$, so $r_q = RR$. $E_q$ is then $\{RR, LR, LL\}$, so $P = \{RRRR, RRLR, RRLL\}$.

By definition, $p \in P$, and $P \subseteq leaves_\sigma(g)$ because $p \in leaves_\sigma(g)$, and $P$ are paths obtained from $\Phi_=(x)$ and $g$ satisfies $\Phi_=(x)$. Moreover, for any term $t$ with the appropriate sort, $g[P \leftarrow t]$ still satisfies $\Phi_=(x)$, because $P$ are all paths equal to $p$ due to $\Phi_=(x)$.

We can also conclude that $g[P \leftarrow t]$ satisfies $\Phi_{=c}(x)$. Let $s(x) = c$ be some atom of $\Phi_{=c}$. As $p \in P$, there is some suffix $q$, s.t. $p = r_q q$ and $q$ is from bidirectional congruence closure graph, so as $\|p\| > K$, $\|p\| = \|r_q\| + \|q\| > K > \|\varphi\| > \|s\|$. Taking any other $r_q e \in P$, s.t. $e \neq q$, we have $\|r_q e\| > K - \|q\| + \|e\| > \|\varphi\| - \|q\| + \|e\| > \|e\| + \|q\| + \|s\| - \|q\| + \|e\| > \|s\|$. Thus, all paths in $P$ are strictly longer than $s$, so $s(g[P \leftarrow t])$ is a leaf only iff $s(g)$ is a leaf, and hence $\Phi_{=c}(x)$ is also satisfied by $g[P \leftarrow t]$ for any $t$.

By the same argument we can conclude that $g[P \leftarrow t]$ satisfies $\Phi_{\neq c}(x), \Phi_{c?}(x)$ and $\Phi_{\neg c?}(x)$. Taking literal $\neg(s(x) = c)$, we again have all paths in $P$ being strictly longer than $s$, so $s(g[P \leftarrow t])$ is a leaf only iff $s(g)$ is a leaf, for any $t$. For atoms $c?(s(x))$ the same applies, and so $s(g[P \leftarrow t])$ has top constructor $c$ only iff $s(g)$ has it.

Finally, it should be shown that $g[P \leftarrow t]$ also satisfies $\Phi_{\neq}(x)$. Unfortunately, this is not true for the arbitrary $t$. Let $\neg(s(x) = s'(x))$ be some literal from the $\Phi_{\neq}(x)$. The fact that $g$ satisfies the literal implies $s(g) \neq s'(g)$. Pumping may fail if, e.g., $s'(g)$ is higher than $s(g)$, the latter is pumped (and becomes equal to the former) and the former is not pumped.

However, if we put $N \stackrel{\text{def}}{=} 1 + \mathcal{H}eight(g)$ and consider only terms $t$ higher than $N$, as the lemma states, pumping works. To make it obvious, that this restriction works, we consider three cases. First, if neither $s$ or $s'$ are suffixes of any path in $P$, then $s(g[P \leftarrow t]) = s(g)$ and $s'(g[P \leftarrow t]) = s'(g)$, so disequality obviously holds. Second, if w.l.o.g. $s$ is a suffix of some path in $P$ and $s'$ is not, $\mathcal{H}eight(s(g[P \leftarrow t])) \geq N > \mathcal{H}eight(g) \geq \mathcal{H}eight(s'(g)) = \mathcal{H}eight(s'(g[P \leftarrow t]))$, so the disequality holds.

The third case, if both $s$ and $s'$ are suffixes of some paths in $P$, is proved by contradiction. Suppose $s(g[P \leftarrow t]) = s'(g[P \leftarrow t])$. As all paths in $P$ are substituted simultaneously and $\mathcal{H}eight(t) > \mathcal{H}eight(g)$, all equal to $t$ subterms of $g[P \leftarrow t]$ are *only* on paths $P$. Thus, it can be shown by induction on the size of $P$, by throwing out elements of $P$, that $s(g) = s'(g)$. The induction step is obvious, as all $t$ subterms are only on paths $P$: for some $r$ and $r'$ if $rs(g[P \leftarrow t]) = r's'(g[P \leftarrow t]) = t$, then $rs, r's' \in P$. Thus, they can be substituted back ($P_1 = P \setminus \{rs, r's'\}$) while the equality still holds: $s(g[P_1 \leftarrow t]) = s'(g[P_1 \leftarrow t])$. By induction, some $P_i$ after the finite number of steps (as $P$ is finite) becomes empty, so $g[P_i \leftarrow t] = g$, and hence $s(g) = s'(g)$, which contradicts the fact that $g$ satisfies the formula. □

**Lemma 6** (Pumping Lemma for Elem). *Let $\mathbf{L}$ be an elementary language of n-tuples. Then, there exists a constant $K > 0$ satisfying: for every n-tuples of ground terms $\langle g_1, \ldots, g_n \rangle \in \mathbf{L}$, for any $i$ such that $\mathcal{H}eight(g_i) > K$, for all infinite sorts $\sigma \in \Sigma_S$ and for all paths $p$ with a length greater than $K$, there exist finite sets of paths $P_j$ such that $p \in P_i$, for all $p_1, p_2 \in \bigcup_j P_j$ it is true that $p_1(g) = p_2(g)$, and there is $N \geq 0$, such that for all $t$ of sort $\sigma$ with $\mathcal{H}eight(t) > N$ it holds that:*

$\langle g_1[P_1 \leftarrow t], \ldots, g_i[P_i \leftarrow t], \ldots, g_n[P_n \leftarrow t] \rangle \in \mathbf{L}.$

*Proof.* Assume that $\mathbf{L}$ is defined by formula $\varphi(x_1, \ldots, x_n)$ with some sorts $x_i : \sigma_i$. We can extend the ADT by adding new sort $\sigma_T$ and fresh symbol $T$ (modeling tuples) with sort $\sigma_1 \times \ldots \times \sigma_n \rightarrow \sigma_T$. In the extended language there is an equisatisfiable to $\varphi$ formula

$$\psi(y) \stackrel{\text{def}}{=} \exists x_1, \ldots, x_n. \, (y = T(x_1, \ldots, x_n) \wedge \varphi(x_1, \ldots, x_n))$$

defining 1-dimensional language $\mathbf{L}'$. That is, for all ground $\langle g_1, \ldots, g_n \rangle$ we have $\langle g_1, \ldots, g_n \rangle \in \mathbf{L}$ iff $T(g_1, \ldots, g_n) \in \mathbf{L}'$. The proof is concluded by applying 1-dimensional pumping lemma. □

From the theory on successor arithmetic it is known that unary successor arithmetic predicates can only define finite unions of finite and cofinite sets [19]. For example, $\varphi(x) \equiv x = 10 \vee x = 15$ defines finite $\{10, 15\}$, $\varphi(x) \equiv \neg(x = 1) \vee x = 5$ defines $\mathbb{N} \setminus \{1\} \cup \{5\}$, etc. Terms of type $Nat = Z | S(Nat)$ are isomorphic to integers, and thus the same property applies to this type. The pumping lemma for ELEM thou can be viewed as *a generalization* of this definability property for *arbitrary* term languages. If we specialize the lemma to the type $Nat$, we get "if big enough $g$ is in the set, then for arbitrary $t > N$, $g + t$ is also in the set", or "every definable set $\mathbf{L}$ is either $\mathbf{L} \subseteq \{x \mid x \leq K\}$ (is finite), or has a subset $\{g+t \mid t > N\} \subseteq \mathbf{L}$ (is cofinite)".

## B.2 Pumping Lemma for SIZEELEM

We incorporate notations of Hojjat and Rümmer [31] and denote $\mathbb{T}^k_\sigma = \{t \text{ has sort } \sigma \mid size(t) = k\}$. For each ADT sort $\sigma$ we define the set of term sizes $\mathbb{S}_\sigma = \{size(t) \mid t \in |\mathcal{H}|_\sigma\}$.

The proof of the pumping lemma for SIZEELEM is based on the concept of a normal form for SIZEELEM formulas.

Let $\mathcal{T}(\varphi(x))$ be a set of paths in testers of formula $\varphi$, i.e., for each $c?(s(x))$ atom in $\varphi$, $s \in \mathcal{T}(\varphi(x))$.

Let $\mathcal{S}(\varphi(x))$ be a set of other paths in formula $\varphi(x)$:

- for each $s(x) = s'(x)$ atom in $\varphi(x)$, $s \in \mathcal{S}(\varphi(x))$ and $s' \in \mathcal{S}(\varphi(x))$;
- for each $size(s(x))$ term in $\varphi(x)$, $s \in \mathcal{S}(\varphi(x))$.

**Definition 7** (Normal Form for SIZEELEM). *A first-order formula with size constraints $\varphi(x)$ of the extended ADT signature is in a normal form iff the following restrictions (1)-(6) hold for $\varphi$.*

**Restriction 1.** *Formula $\varphi$ has the form:*

$$\varphi(x) \equiv \bigvee_i \left( \Phi_i^{Elem}(x) \wedge \Phi_i^{Size}(x) \right),$$

*where each $\Phi_i^{Elem}$ is a conjunction of literals in one of the forms:*

- *$c?(s(x))$, for some tester $c?$ and some path $s$,*
- *$s(x) = s'(x)$, for some paths $s, s'$,*
- *$\neg(s(x) = s'(x))$, for some paths $s, s'$,*

*and each $\Phi_i^{Size}$ has no constructors, testers and no equalities between non-integer terms, so it should be representable as*

$$\Phi_i^{Size}(x) \equiv \Phi_i^{Pres}(size(s_1(x)), \ldots, size(s_m(x))),$$

*where $\Phi_i^{Pres}(n_1, \ldots, n_m)$ is some Presburger formula defining a linear set. Each conjunct in $\varphi$ should not be equivalent to $\bot$.*

**Restriction 2.** *For each $i$, if $(s(x) = s'(x)) \in \Phi_i^{Elem}$ then $(size(s(x)) = size(s'(x))) \in \Phi_i^{Size}$.*

**Restriction 3.** *For each $i$, each $size(s(x))$ subterm in $\Phi_i^{Size}$, where $s(x)$ has some sort $\sigma$, and each integer constant $c$, if $\Phi_i^{Size} \wedge size(s(x)) = c$ is satisfiable, then there must exist some ground $\sigma$-term $g$ with $size(g) = c$.*

**Restriction 4.** *For each $i$, $\mathcal{S}(\Phi_i^{Elem}) \subseteq \mathcal{S}(\Phi_i^{Size})$.*

**Restriction 5.** *For each $i$, for any paths $q, r \in \mathcal{S}\left(\Phi_i^{Size}\right)$ if $q \neq r$, then neither $q$ or $r$ should be suffixes one of another.*

**Restriction 6.** *For each $i$, the following should hold:*

$$\mathcal{T}\left(\Phi_i^{Elem}\right) = \left\{q \mid \exists r \in \mathcal{S}\left(\Phi_i^{Size}\right), q \text{ is a proper suffix of } r\right\}.$$

**Lemma 9.** *Each SIZEELEM formula has an equivalent formula in a normal form.*

*Proof.* It was shown by Zhang et al. [58] that SIZEELEM admits quantifier elimination, so we can eliminate quantifiers, translate the formula to DNF and perform a number of rewritings until none of them is applicable. If new disjunctions under conjunctions appear, the formula is again getting transformed into DNF. To avoid unnecessary complexity while defining rewritings, we present them on LISP-style lists — the approach can be easily generalized for any ADTs. Each rewriting is defined by $a \leadsto b$ patterns, meaning the replacement of term $a$ with $b$ or subformula $a$ with $b$. Also after each rewriting occurs-check and standard simplifications are applied, e.g., $x = x$ is replaced with $\top$, $car(x) = x$ is replaced with $\bot$, $\varphi \wedge \bot$ is replaced with $\bot$, etc.

**Rewriting Rule 1.** First, we can remove all constructors by introducing new selectors and testers:

$$size(cons(a, b)) \leadsto 1 + size(a) + size(b)$$
$$car(cons(a, b)) \leadsto a$$
$$cons(a, b) = x \leadsto cons?(x) \wedge car(x) = a \wedge cdr(x) = b.$$

Second, all testers in negative context can be substituted with disjunction of positive testers:

$$\neg nil?(x) \leadsto \bigvee_{c \neq nil} c?(x) \leadsto cons?(x).$$

All equalities with constant constructor can be substituted with appropriate testers:

$$t = nil \leadsto nil?(t).$$

If $\Phi_i^{Size}$ is built from $\Phi_i^{Pres}$ which defines a *semi*-linear set $\bigcup_j L_j$ for linear sets $L_j$, $\Phi_i^{Pres}$ and thus $\Phi_i^{Size}$ can be represented as an equivalent disjunction $\bigvee_j \Psi_j$, where each $\Psi_j$ defines a linear set $L_j$. These rewritings already fulfill the restriction 1.

**Rewriting Rule 2.** For each $i$, for each literal $s(x) = s'(x)$ from $\Phi_i^{Elem}$ we conjunct the literal $size(s(x)) = size(s'(x))$ to the formula $\Phi_i^{Size}$, which fulfills the restriction 2.

**Rewriting Rule 3.** As it was stated by Hojjat and Rümmer [31], the size image is a special case of the Parikh image of the ADT declaration, viewed as a context-free grammar over the singleton alphabet (by interpreting every ADT sort as a non-terminal, and mapping every constructor to the only letter in alphabet). By Parikh's theorem, $\mathbb{S}_\sigma$ is semi-linear, so it can be represented by some Presburger formula $\psi_\sigma(n)$. By its definition, for each $n_0$ satisfying $\psi_\sigma$ there exists some ground $\sigma$-term $g$, s.t. $size(g) = n_0$. That is, we can fulfill the restriction 3 by conjuncting to each $\Phi_i^{Size}$ all formulas $\psi_\sigma(size(s(x)))$ for each paths $s \in \mathcal{T}(\Phi_i^{Elem}) \cup \mathcal{S}(\Phi_i^{Elem} \wedge \Phi_i^{Size})$ with $s(x)$ of sort $\sigma$. After such an extension we have
$$\mathcal{S}(\Phi_i^{Elem}) \subseteq (\Phi_i^{Size}),$$
and hence the restriction 4 is also fulfilled. We also have
$$\mathcal{T}(\Phi_i^{Elem}) \subseteq \mathcal{S}(\Phi_i^{Size}). \quad (1)$$

**Rewriting Rule 4.** If restriction 5 is not fulfilled, there are some paths, which are suffixes of one another, e.g., $car(cdr(x))$ and $cdr(x)$. The shorter one can be eliminated by the rule:
$$\psi(cdr(x)) \rightsquigarrow \psi(cons(car(cdr(x)), cdr(cdr(x)))) \wedge \quad (2)$$
$$\wedge cons?(cdr(x)).$$
Introduced constructors are eliminated by the previous rules and short paths like $cdr$ are replaced with longer $carcdr$ and $cdrcdr$. This rewriting rule must be applied only to non-tester atoms (to not violate the restriction 6), and it allows us to fulfill the restriction 5.

**Rewriting Rule 5.** By (1) it is true that for each $q \in \mathcal{T}(\Phi_i^{Elem})$ there exists $r \in \mathcal{S}(\Phi_i^{Elem} \wedge \Phi_i^{Size})$, s.t. $q = r$. In order to fulfill the restriction 6, we can apply the rule (2) to all non-tester atoms with left-hand side of the rule matching only elements of $\mathcal{T}(\Phi_i^{Elem})$.

All rules produce equivalent formulas, so the final formula is in a normal form and is equivalent to the starting formula. □

**Example 6.** Consider the ADT type
$$Tree = leaf \mid node(L : Tree, R : Tree).$$
Let us denote $|t|$ a $size(t)$ for brevity. Consider formula $\varphi_0(x) \equiv \exists l, r, ll.(x = node(l, r) \wedge l = node(ll, r) \wedge |ll| < |r|)$. We can find an equivalent formula in a normal form by applying the above lemma. We denote by $\rightsquigarrow_i$ application of the rewriting rule with number $i$. First, we eliminate quantifiers and constructors:
$$\varphi_0(x) \rightsquigarrow_1 \varphi_1^{Elem} \wedge \varphi_1^{Size},$$
where $\varphi_1^{Elem} \equiv node?(x) \wedge node?(Lx) \wedge RLx = Rx$ and $\varphi_1^{Size} \equiv |LLx| < |Rx|$. Then, we introduce the implied equalities of sizes:
$$\varphi_1^{Elem} \wedge \varphi_1^{Size} \rightsquigarrow_2 \varphi_1^{Elem} \wedge \varphi_2^{Size},$$
where $\varphi_2^{Size} \equiv \varphi_1^{Size} \wedge |RLx| = |Rx|$. After that, ADT definition implied restrictions on sizes are added:
$$\varphi_1^{Elem} \wedge \varphi_2^{Size} \rightsquigarrow_3 \varphi_1^{Elem} \wedge \varphi_3^{Size},$$
where $\varphi_3^{Size} \equiv \varphi_2^{Size} \wedge \xi_1 \wedge \xi_2$,
$\xi_1 \equiv 2 \mid (|Rx| + 1) \wedge 2 \mid (|LLx| + 1) \wedge 2 \mid (|RLx| + 1)$
$\xi_2 \equiv 2 \mid (|x| + 1) \wedge 2 \mid (|Lx| + 1)$,

as sizes of possible tree terms are odd numbers $\mathbb{S}_{Tree} = \{1, 3, 5, \ldots\}$ which is represented via[12] $\psi_{Tree}(n) \equiv 2 \mid (n + 1)$.
$\xi_2 \rightsquigarrow_4 2 \mid (|node(Lx, Rx)| + 1) \wedge 2 \mid (|node(LLx, RLx)| + 1)$
$\rightsquigarrow_1 2 \mid (|Lx| + |Rx| + 2) \wedge 2 \mid (|LLx| + |RLx| + 2)$
$\sim 2 \mid (|Lx| + |Rx|) \wedge 2 \mid (|LLx| + |RLx|)$
$\rightsquigarrow_4 2 \mid (|node(LLx, RLx)| + |Rx|) \wedge 2 \mid (|LLx| + |RLx|)$
$\rightsquigarrow_1 2 \mid (|LLx| + |RLx| + 1 + |Rx|) \wedge 2 \mid (|LLx| + |RLx|)$
$\sim 2 \mid (1 + |Rx|) \wedge 2 \mid (|LLx| + |RLx|) \equiv \xi_3$
It is easy to see that $\xi_1 \models \xi_3$, so
$$\varphi_1^{Elem} \wedge \varphi_3^{Size} \rightsquigarrow^* \varphi_1^{Elem} \wedge \varphi_4^{Size},$$
where $\varphi_4^{Size} \equiv \varphi_2^{Size} \wedge \xi_1$. So we finally get a formula
$$node?(x) \wedge node?(Lx) \wedge RLx = Rx \wedge$$
$$\wedge |LLx| < |Rx| \wedge |RLx| = |Rx| \wedge 2 \mid (|Rx| + 1) \wedge$$
$$\wedge 2 \mid (|LLx| + 1) \wedge 2 \mid (|RLx| + 1), \quad (3)$$
which is equivalent to the starting formula $\varphi_0(x)$ and is in the normal form.

For pumping lemma we should offer a suitable lower bound $K$ of term height, so that the pumping could be successful. For SizeElem class it is not enough to take the size of the formula (as we did for the Elem class), because size constraints can limit the height of terms as well (e.g., a formula $size(x) < 100$ defines a finite language, which is obviously not pumpable). By definition of the SizeElem language, each size constraining part of the SizeElem formula can be represented as Presburger formula over sizes
$$\Phi^{Size} \equiv \Phi^{Pres}(size(s_1(x)), \ldots, size(s_m(x))).$$

**Definition 8.** Let us define a *finite maximum* of Presburger formula $\Phi(n_1, \ldots, n_d)$, denoted as $\max_{fin}(\Phi)$. $\Phi$ defines some semi-linear set $\bigcup_{j=1}^{l} L_j$ for some finite $l$, where
$$L_j = \left\{ \mathbf{v}_0^{(j)} + \sum_{i=1}^{i=m_j} k_i \mathbf{v}_i^{(j)} \mid k_1, \ldots, k_{m_j} \in \mathbb{N}_0 \right\},$$
where each $\mathbf{v}_i^{(j)}$ is a vector of natural numbers of dimension $d$.
$$\max_{fin}(\Phi) \stackrel{\text{def}}{=} \max_{j=1}^{l} \max_{a=1}^{d} \begin{cases} \mathbf{v}_0^{(j)}[a], & \text{if } \forall 1 \leq i \leq m_j, \mathbf{v}_i^{(j)}[a] = 0 \\ 0, & \text{otherwise} \end{cases}$$
where $\mathbf{v}[a]$ is the $a$-th component of the vector $\mathbf{v}$.

Observe that if $\max_{fin}(\Phi) = 0$, then $\Phi$ is either unsatisfiable, or satisfied by infinite number of integers.

If $\varphi(x)$ is a SizeElem formula in a normal form,
$$\varphi(x) \equiv \bigvee_i \left( \Phi_i^{Elem}(x) \wedge \Phi_i^{Size}(x) \right),$$
and for each $i$, $\Phi_i^{Size}(x) \equiv \Phi_i^{Pres}(t_1, \ldots, t_{n_i})$, where each $\Phi_i^{Pres}$ is a Presburger formula, define
$$\max_{fin}(\varphi) \stackrel{\text{def}}{=} \max_i \max_{fin}(\Phi_i^{Pres}).$$

To avoid other problems with pumping, we pump only *expanding* ADTs. Here we restate the definition of expanding ADTs from Sec. 6.3, which we borrow from [31].

---
[12]"$2 \mid x$" means "2 divides $x$", i.e., $x$ is an even integer.

**Definition 9.** An ADT sort $\sigma$ is *expanding* iff for every natural number $n$ there is a bound $b(\sigma, n) \geq 0$ such that for each $b' \geq b(\sigma, n)$, if $\mathbb{T}_\sigma^{b'} \neq \emptyset$, then $|\mathbb{T}_\sigma^{b'}| \geq n$. An ADT signature is called expanding if all it's sorts are expanding.

**Example 7.** Consider ADTs:
$$Nat ::= Z \mid S(Nat)$$
$$List ::= nil \mid cons(Nat, List)$$

For each $k$, $|\mathbb{T}_{Nat}^k| = 1$, so the $Nat$ sort is not expanding.

Starting with $k = 3$, $|\mathbb{T}_{List}^k| = fib(k - 2)$ and for $i \geq 5$, $fib(i) \geq i$, so taking $b = n + 7$,
$$|\mathbb{T}_{List}^{b'}| > |\mathbb{T}_{List}^b| = fib(b - 2) = fib(n + 5) \geq n,$$
and thus the $List$ sort is expanding.

Expansive ADTs are sufficiently populated to satisfy disequalities in a formula. For example, $\neg(x = y) \wedge size(x) = size(y)$ is unsatisfiable if $x$ and $y$ have non-expanding sort $Nat$ and is satisfiable if they have expanding sort $List$.

**Lemma 10.** *If $A$ and $B$ are infinite 1-dimensional linear sets, $A \cap B$ is either empty or it is also an infinite linear set.*

*Proof.* Let
$$A = \{a \mid a = w_0 + \sum_i k_i w_i, k_i \in \mathbb{N}_0\}$$
$$B = \{b \mid b = v_0 + \sum_j k_j v_j, k_j \in \mathbb{N}_0\}$$
$$A \cap B \ni c = w_0 + \sum_i k_i^1 w_i = v_0 + \sum_j k_j^2 v_j,$$

for some $k_i^1, k_j^2$. As both $A$ and $B$ are infinite, $W \stackrel{\text{def}}{=} \sum_i w_i > 0$ and $V \stackrel{\text{def}}{=} \sum_j v_j > 0$. Consider
$$c + dWV = w_0 + \sum_i k_i^1 w_i + dV \sum_i w_i =$$
$$= w_0 + \sum_i (k_i^1 + dV) w_i \in A,$$
$$c + dWV = v_0 + \sum_j k_j^2 v_j + dW \sum_j v_j =$$
$$= v_0 + \sum_j (k_j^2 + dW) v_j \in B.$$

Hence for arbitrary $d \in \mathbb{N}_0$, $c + dWV \in A \cap B$ and thus $A \cap B$ is infinite. Moreover, it is linear [27]. $\square$

We are now ready to state and prove the pumping lemma for one-dimensional languages of the SizeElem class.

**Lemma 11** (1-dimensional pumping lemma for SizeElem). *Let the ADT signature be expanding and let $\mathbf{L}$ be a 1-dimensional elementary language with size constraints. Then, there exists a constant $K > 0$ satisfying: for every ground term $g \in \mathbf{L}$ such that $\mathcal{H}eight(g) > K$, for all infinite sorts $\sigma \in \Sigma_S$ and for all paths $p \in leaves_\sigma(g)$ with $\|p\| > K$, there exists an infinite linear set $T \subseteq \mathbb{S}_\sigma$, such that for all terms $t$ of sort $\sigma$ with sizes $size(t) \in T$, there exists a sequence of paths $P$, none of which is a suffix of the path $p$, and a sequence of terms $U$, such that $g[p \leftarrow t, P \leftarrow U] \in \mathbf{L}$.*

*Proof.* By definition, $\mathbf{L}$ is defined by some SizeElem formula $\varphi(x)$. By Lemma 9 we can assume that $\varphi(x)$ is in a normal form. We put
$$K \stackrel{\text{def}}{=} \max_\sigma\{size(t) \mid t \text{ is a leaf term of ADT sort } \sigma\} +$$
$$+ \|\varphi\| + \max_{fin}(\varphi).$$

Let $g \in \mathbf{L}$ be such that $\mathcal{H}eight(g) > K$, let $\sigma \in \Sigma_S$ be an infinite sort, and let $p \in leaves_\sigma(g)$ be a path with $\|p\| > K$. As $g \in \mathbf{L}$, $g$ satisfies some conjunct in
$$\varphi(x) \equiv \bigvee_i \left( \Phi_i^{Elem}(x) \wedge \Phi_i^{Size}(x) \right),$$
let us denote it $\Phi(x) \equiv \Phi^{Elem}(x) \wedge \Phi^{Size}(x)$.

Now we can pick some path $q \in \mathcal{S}(\Phi^{Size})$, which is a suffix of $p$. Consider two cases. If $q$ does not exist $(\mathcal{S}(\Phi^{Size}) = \emptyset)$, by the definition of the $\mathcal{S}$, the formula has no constraints on path $p$, so the subterm $p(g)$ at this path in $g$ can be substituted with any $t$ of sort $\sigma$. That is, if we put $T \stackrel{\text{def}}{=} \mathbb{S}_\sigma$, we obtain an infinite (as $\sigma$ is an infinite sort) semi-linear set and can take any of it's infinite linear subsets as $T$. We can then take $P = U = \emptyset$, so for any $t$ of sort $\sigma$ with $size(t) \in T$, $g[p \leftarrow t, P \leftarrow U] = g[p \leftarrow t] \in \mathbf{L}$, as the subterm $p(g)$ is not constrained by formula and thus can be replaced with the term $t$, still satisfying the formula.

The second case, when there exists some path $q$, which is a suffix of $p$, is more complex. By the restriction 5, if it exists, it is unique. By the definition of the SizeElem class and by the restriction 1,
$$\Phi^{Size}(x) \equiv \Phi^{Pres}(size(s_1(x)), \ldots, size(s_m(x))),$$
for some formula of Presburger arithmetic $\Phi^{Pres}(a_1, \ldots, a_m)$. W.l.o.g., let us assume that $q = s_1$. We define $P$ to be the rest of the paths: $P \stackrel{\text{def}}{=} \langle s_2; \ldots; s_m \rangle$. As $q$ is the only suffix of $p$, none of the paths of $P$ are suffixes of $p$.

To proceed further, we have to take a detailed look on $\Phi^{Pres}$, which by restriction 1 defines some linear set
$$L = \left\{ \mathbf{v}_0 + \sum_{i=1}^{i=l} k_i \mathbf{v}_i \mid k_1, \ldots, k_l \in \mathbb{N}_0 \right\},$$
where all $\mathbf{v}_j \in \mathbb{N}_0^m$. We define
$$L(\overline{k}) = \mathbf{v}_0 + \sum_{i=1}^{i=l} k_i \mathbf{v}_i,$$
so
$$L(\overline{k})[1] = \mathbf{v}_0[1] + \sum_{i=1}^{i=l} k_i \mathbf{v}_i[1]$$
is the constraint on the subterms in path $q$. Note that
$$size(q(g[p \leftarrow t])) = size(q(g)) - size(p(g)) + size(t),$$
for arbitrary $t$. We denote $c = size(p(g))$, as it is constant. There is some $\overline{k^0}$, such that $size(q(g)) = L(\overline{k^0})[1]$, and we can choose arbitrary $\overline{k}$, such that $size(q(g[p \leftarrow t])) =$

$L(\overline{k})[1]$, satisfying
$$L(\overline{k})[1] = L(\overline{k^0})[1] - c + size(t)$$
$$\mathbf{v}_0[1] + \sum_{i=1}^{i=l} k_i \mathbf{v}_i[1] = \mathbf{v}_0[1] + \sum_{i=1}^{i=l} k_i^0 \mathbf{v}_i[1] - c + size(t)$$
$$size(t) = c + \sum_{i=1}^{i=l} (k_i - k_i^0) \mathbf{v}_i[1].$$

Now if we put $A_i \stackrel{def}{=} k_i - k_i^0 \geq 0$, then
$$size(t) = c + \sum_{i=1}^{i=l} A_i \mathbf{v}_i[1],$$
for any $A_i \in \mathbb{N}_0$. From the above equation we can derive restrictions on $t$, which will be included in the set $T$ from the lemma statement. Let $\mathbf{v} = \langle \mathbf{v}_1, \ldots, \mathbf{v}_l \rangle$ and $\mathbf{v}[j] \stackrel{def}{=} \langle \mathbf{v}_1[j], \ldots, \mathbf{v}_l[j] \rangle$ for all $j$. We can define
$$T_1 \stackrel{def}{=} \left\{ c + \sum_{i=1}^{i=l} A_i \mathbf{v}_i[1], A_i \in \mathbb{N}_0 \right\}.$$

$T_1$ is infinite because $\mathbf{v}[1] \neq \overline{0}$, which can be proven by contradiction: if $\mathbf{v}[1] = \overline{0}$ then $\max_{fin} (\varphi) \geq \mathbf{v}_0[1]$, so
$$\|p\| > K > \|\varphi\| + \max_{fin} (\varphi) \geq \|q\| + \mathbf{v}_0[1],$$
so $\|p\| - \|q\| > \mathbf{v}_0[1]$. On the other hand,
$$size(q(g)) \geq (size(p(g)) - 1) + (\|p\| - \|q\|) > \mathbf{v}_0[1],$$
but $size(q(g))$ must satisfy $\Phi^{Pres}$ and be equal to $\mathbf{v}_0[1]$, as $g$ satisfies $\Phi^{Size}$, contradiction. So $T_1$ is an infinite linear set.

Let us partition all paths from the formula into groups:
$$\mathcal{C}_\sigma \stackrel{def}{=} \{s \in \mathcal{S}(\Phi^{Size}) \mid sx \text{ has sort } \sigma\}/\sim,$$
where $s \sim r$ is a reflexive, symmetric and transitive closure of the relation
$$s \sim' r \Leftrightarrow (sx = rx) \in \Phi^{Elem}.$$

We are now ready to construct $U = \langle u_2, \ldots, u_m \rangle$ — terms which will be placed at paths $P = \langle s_2, \ldots, s_m \rangle$. For each $2 \leq j \leq m$, with $s_j(x)$ of sort $\sigma_j$ we have a constraint on $s_j(x)$:
$$L(\overline{k})[j] = \mathbf{v}_0[j] + \sum_{i=1}^{i=l} k_i \mathbf{v}_i[j].$$

If $\mathbf{v}[j] = \overline{0}$, then $u_j \stackrel{def}{=} s_j(g)$. Otherwise, there always can be chosen such $\overline{d_j} \in \mathbb{N}_0^l$, that for all $\overline{k}$, s.t. for all $i$, $k_i \geq \overline{d_j}_i$:
$$L(\overline{k})[j] \geq b(\sigma_j, |\mathcal{C}_{\sigma_j}|),$$
where $b(\sigma_j, n)$ is a bound for $\sigma_j$ from the definition of the expanding ADT with sort $\sigma_j$. As $\overline{k} = \overline{k^0} + \overline{A}$, we must have for all $i$, $A_i \geq \max(0, \overline{d_j}_i - k_i^0)$. Take $E_i \stackrel{def}{=} \max_{j=2}^{m} \max(0, \overline{d_j}_i - k_i^0)$, so we can define:
$$T_2 \stackrel{def}{=} \left\{ c + \sum_{i=1}^{i=l} (A_i + E_i) \mathbf{v}_i[1], A_i \in \mathbb{N}_0 \right\} =$$
$$= \left\{ \left( c + \sum_{i=1}^{i=l} E_i \mathbf{v}_i[1] \right) + \sum_{i=1}^{i=l} A_i \mathbf{v}_i[1], A_i \in \mathbb{N}_0 \right\}.$$

$T_2$ is again, obviously infinite and linear. It has all necessary properties: for all terms $t$ with sort $\sigma$ and $size(t) \in T_2$, $size(t)$ fixes some $\overline{A}$, s.t. for all $i$, $A_i \geq \max(0, \overline{d_j}_i - k_i^0)$ and hence $k_i = k_i^0 + A_i \geq \overline{d_j}_i$, so $c_j \stackrel{def}{=} L(\overline{k})[j] \geq b(\sigma_j, |\mathcal{C}_{\sigma_j}|)$. By the restriction 3, $\mathbb{T}_{\sigma_j}^{c_j} \neq \emptyset$, so by definition of the expanding ADT we finally have
$$\left| \mathbb{T}_{\sigma_j}^{c_j} \right| \geq \left| \mathcal{C}_{\sigma_j} \right|.$$

Now for each sort $\sigma_i$ we can take $|\mathcal{C}_{\sigma_i}|$ distinct terms $\overline{t^i}$ from $\mathbb{T}_{\sigma_i}^{c_i}$. For each $j, j'$, if $s_j(x)$ and $s_{j'}(x)$ have sort $\sigma_i$, then if $[s_j] = [s_{j'}] \in \mathcal{C}_{\sigma_i}$ then $u_j \stackrel{def}{=} u_{j'} \stackrel{def}{=} t \in \overline{t^i}$. Otherwise, if $[s_j] \neq [s_{j'}]$ then $u_j \stackrel{def}{=} t \in \overline{t^i}$ and $u_{j'} \stackrel{def}{=} t' \in \overline{t^i}$ for some $t \neq t'$.

As $\sigma$ is infinite, $\mathbb{S}_\sigma$ is also infinite. As $size(p(g)) \in T_1 \cap \mathbb{S}_\sigma$, we conclude by Lemma 10 that this intersection is an infinite linear set. It follows from definitions that $T_2 \subseteq T_1$, so $T \stackrel{def}{=} T_1 \cap \mathbb{S}_\sigma \cap T_2$ is also an infinite linear set.

The last thing to show is that $g[p \leftarrow t, P \leftarrow U]$ satisfies $\Phi^{Elem}(x) \wedge \Phi^{Size}(x)$.

It satisfies $\Phi^{Size}(x)$ because $\{q\} \cup P = \mathcal{S}(\Phi^{Size})$ and each $u_j$ is either $s_j(g)$ (and $g$ satisfies $\Phi^{Size}(x)$), or some $t \in \mathbb{T}_{\sigma_j}^{c_j}$, and $size(u_j) = size(t) = c_j = L(\overline{k})[j]$ which is precisely the size constraint from $\Phi^{Size}(x)$.

It satisfies $\Phi^{Elem}(x)$ because all its subterms are determined by $P$ (follows from the restriction 4) and the choice of $u_j$ preserves the equality classes of $\mathcal{C}_{\sigma_j}$ (which preserve sizes by the restriction 2), which is a factorization by equalities obtained from $\Phi^{Elem}(x)$. As $u_j$ for paths from distinct equality classes are distinct, all disequalities from $\Phi^{Elem}(x)$ are again automatically satisfied. Finally, all testers from $\Phi^{Elem}(x)$ are satisfied as their paths are *proper* suffixes of paths in $\mathcal{S}(\Phi^{Size})$ (by the restriction 6), so the top constructor is untouched. □

The restriction of the ADT to be expanding is substantial and also used in the satisfiability checking procedure in [31]. However, we have more opportunities to weaken this restriction and still overcome the inhabitation problem in the pumping lemma. Intuitively, we can make $K$, the height bound of $g$, as large as possible. Term $g \in \mathbf{L}$ can be viewed as a witness that for big enough terms there are enough distinct ground terms with certain size to satisfy all disequalities. We leave weakening of expansibility restriction for the future work.

Just as for Elem, general lemma for SizeElem can be proved by applying 1-dimensional lemma.

**Lemma 7** (Pumping Lemma for SizeElem). *Let the ADT signature be expanding and let $\mathbf{L}$ be an elementary language of n-tuples with size constraints. Then, there exists a constant $K > 0$ satisfying: for every n-tuple of ground terms $\langle g_1, \ldots, g_n \rangle \in \mathbf{L}$, for any $i$, such that $\mathcal{H}eight(g_i) > K$, for all infinite sorts $\sigma \in \Sigma_S$, and for all paths $p \in leaves_\sigma(g_i)$ with length greater than $K$, there exists an infinite linear set $T \subseteq \mathbb{S}_\sigma$, such that for*

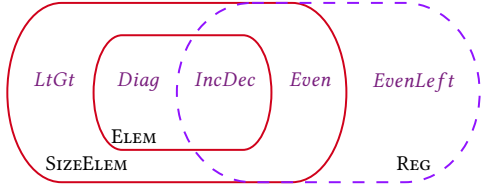

**Figure 3.** The comparison of the expressive power of three computable representations of inductive invariants.

*all terms $t$ of sort $\sigma$ with sizes $size(t) \in T$, there exist sequences of paths $P_j$, with no path in them being a suffix of path $p$, and sequences of terms $U_j$, such that*

$$\langle g_1[P_1 \leftarrow U_1], \ldots, g_i[p \leftarrow t, P_i \leftarrow U_i], \ldots, g_n[P_n \leftarrow U_n]\rangle \in \mathbf{L}.$$

*Proof.* Follows from one-dimensional lemma in the same way as in the proof of Lemma 6. □

## C Comparison of the Expressive Power

The results of comparing the expressiveness of three representations in this section are summarized in Figure 3.

**Example 8** (*IncDec*). Recall the *IncDec* example:

$inc(x, y) \leftarrow x = Z \wedge y = S(Z)$
$inc(x, y) \leftarrow x = S(x') \wedge y = S(y') \wedge inc(x', y')$
$dec(x, y) \leftarrow x = S(Z) \wedge y = Z$
$dec(x, y) \leftarrow x = S(x') \wedge y = S(y') \wedge dec(x', y')$
$\bot \leftarrow inc(x, y) \wedge dec(x, y).$

The least fixed point of *inc* and *dec* are non-regular, i.e., there is no 2-automaton accepting one of these relations [14]. However, this program has another, less obvious regular invariant.

**Proposition 4.** *IncDec* ∈ Reg.

*Proof.* The invariant is induced by two 2-DFTA
$$(\{s_0, s_1, s_2, s_3\}, \Sigma_F, Q, \Delta)$$
with final states correspondingly $Q_{inc} = \{\langle s_0, s_1\rangle, \langle s_1, s_2\rangle, \langle s_2, s_0\rangle\}$, $Q_{dec} = \{\langle s_1, s_0\rangle, \langle s_2, s_1\rangle, \langle s_0, s_2\rangle\}$ and with the following transition rules:

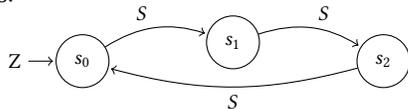

Informally, the automaton for *inc* checks that
$$(x \bmod 3, y \bmod 3) \in \{(0, 1), (1, 2), (2, 0)\},$$
and the automaton for *dec* checks
$$(x \bmod 3, y \bmod 3) \in \{(1, 0), (2, 1), (0, 2)\}.$$

These relations over-approximate the strongest invariants of *inc* and *dec*, but still prove the unsatisfiability of $inc(x, y) \wedge dec(x, y)$. □

That is, *IncDec* ∈ Elem ⊆ SizeElem and also *IncDec* ∈ Reg, so the intersection of these classes is not empty.

**Proposition 5.** Elem ∩ Reg ≠ ∅.

*Proof.* Follows from Example 4 and Prop. 4. □

**Example 9** (*Even*). Recall the Example 1:

$even(x) \leftarrow x = Z$
$even(x) \leftarrow x = S(S(y)) \wedge even(y)$
$\bot \leftarrow even(x) \wedge even(y) \wedge y = S(x)$

**Proposition 6.** *Even* ∈ Reg.

*Proof.* It has a regular model $\langle \mathcal{H}, X\rangle$ induced by 1-DFTA $(\{s_0, s_1, s_2\}, \Sigma_F, \{s_0\}, \Delta)$ where $\Delta$ is the following set of transition rules:

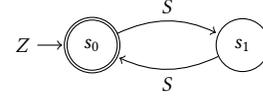

□

Using the pumping lemma for Elem, we have shown in Prop. 1 that *Even* has no elementary invariant.

**Proposition 7.** Reg \ Elem ≠ ∅.

*Proof.* Follows from Prop. 6 and Prop. 1. □

However, the size constraints allow to represent the invariant of the *Even* example:

**Proposition 8.** *Even* ∈ SizeElem.

*Proof.* *Even* has the safe inductive invariant defined by
$$even(x) \Leftrightarrow \exists n. size(x) = 1 + n + n.$$
□

A slight modification of the above *Even* example gives us the problem *EvenLeft* with no invariant, definable by SizeElem.

**Example 10** (*EvenLeft*). Recall the *EvenLeft* example:

$EvenLeft(x) \leftarrow x = leaf$
$EvenLeft(x) \leftarrow x = node(node(x', y), z) \wedge EvenLeft(x')$
$\bot \leftarrow EvenLeft(x) \wedge EvenLeft(node(x, y))$

In Prop. 2 we have shown that *EvenLeft* ∉ SizeElem. However, it still has the regular safe inductive invariant.

**Proposition 9.** *EvenLeft* ∈ Reg.

*Proof.* It is justified by a tree automaton $(\{s_0, s_1, s_2\}, \Sigma_F, \{s_0\}, \Delta)$ with $\Delta$:

$leaf \mapsto s_0$
$node(s_0, s_0) \mapsto s_1$
$node(s_0, s_1) \mapsto s_1$
$node(s_1, s_0) \mapsto s_0$
$node(s_1, s_1) \mapsto s_0$

□

That is, *EvenLeft* ∈ Reg but *EvenLeft* ∉ SizeElem.

**Proposition 10.** Reg \ SizeElem ≠ ∅.

*Proof.* Follows from Prop. 9 and Prop. 2. □

As we see, in lots of cases regular languages are more expressive than the elementary ones. However, it turns out that there are programs with elementary, but not regular invariants.

**Example 11** (*Diag*). Consider the CHC-system obtained from the program over the datatype $Nat := Z : Nat \mid S : Nat \rightarrow Nat$ with the procedure *diseq* recursively checking the disequality of two integer terms.

$$eq(x, y) \leftarrow x = y$$
$$diseq(x, y) \leftarrow x = S(x') \land y = Z$$
$$diseq(x, y) \leftarrow y = S(y') \land x = Z$$
$$diseq(x, y) \leftarrow x = S(x') \land y = S(y') \land diseq(x', y')$$
$$\bot \leftarrow eq(x, y) \land diseq(x, y)$$

Despite the fact that regular languages are closed under intersection, union and complement, tree automata cannot represent the disequality relation. The main intuition behind that is that tree automata can only keep a finite amount of information in states, and they do not precisely relate components of tuples for tuple languages.

**Proposition 11.** $\text{Elem} \setminus \text{Reg} \neq \emptyset$.

*Proof.* It was shown in Comon et al. [14] that $Diag \notin \text{Reg}$.

Although, $Diag \in \text{Elem}$ because of the invariant $I(x, y) \Leftrightarrow x = y$, $diseq(x, y) \Leftrightarrow \neg(x = y)$. □

Size constraints also allow us to define non-regular (and non-elementary) languages.

**Example 12** (*LtGt*). Consider the program *LtGt* recursively comparing the Peano numbers with the following CHC-system:

$$lt(x, y) \leftarrow x = Z \land y = S(y')$$
$$lt(x, y) \leftarrow x = S(x') \land y = S(y') \land lt(x', y')$$
$$gt(x, y) \leftarrow x = S(x') \land y = Z$$
$$gt(x, y) \leftarrow x = S(x') \land y = S(y') \land gt(x', y')$$
$$\bot \leftarrow lt(x, y) \land gt(x, y)$$

**Proposition 12.** $\text{SizeElem} \setminus \text{Reg} \neq \emptyset$.

*Proof.* First, $LtGt \in \text{SizeElem}$ because it has the following invariant, which can be defined by the SizeElem formula:

$$lt(x, y) \Leftrightarrow size(x) < size(y)$$
$$gt(x, y) \Leftrightarrow size(x) > size(y).$$

Now we need to show that ordering is undefinable in Reg. Assume *lt* and *gt* have corresponding invariants: regular languages $L$ and $G$. As regular languages are closed under union, $L \cup G = Diag = \{(x, y) \mid x \neq y\}$ must also be regular, but it is not (see Prop. 11).

□